\documentclass[pra,aps,twocolumn,superscriptaddress]{revtex4}

\usepackage{graphicx}
\usepackage{epsfig}
\usepackage{times}
\usepackage{amsmath}
\usepackage{amsfonts}
\usepackage{amssymb}
\usepackage{latexsym}
\usepackage[usenames]{color}
\usepackage{graphicx}
\usepackage{enumerate}
 \usepackage{times}
 \usepackage{algorithm}
 \usepackage{algorithmic}
 \usepackage{cases}

\newcommand{\black}{\color{black}}
\newcommand{\red}{\color{black}}
\newcommand{\blue}{\color{black}}
\newtheorem{theorem}{Theorem}
\newtheorem{lemma}{Lemma}

\newtheorem{corollary}{Corollary}
\newcommand{\redd}{\color{black}}

\begin{document}
\widetext
\title{Quantum remote sensing with asymmetric information gain}
\author{Yuki Takeuchi}
\email{takeuchi.yuki@lab.ntt.co.jp}
\affiliation{NTT Communication Science Laboratories, NTT Corporation, 3-1 Morinosato-Wakamiya, Atsugi, Kanagawa 243-0198, Japan}
\author{Yuichiro Matsuzaki}
\email{matsuzaki.yuichiro@lab.ntt.co.jp}
\affiliation{NTT Basic Research Laboratories, NTT Corporation, 3-1 Morinosato-Wakamiya, Atsugi, Kanagawa 243-0198, Japan}
\affiliation{NTT Theoretical Quantum Physics Center, NTT Corporation, 3-1 Morinosato-Wakamiya, Atsugi, Kanagawa 243-0198, Japan}
\author{Koichiro Miyanishi}
\affiliation{Graduate School of Engineering Science, Osaka University,
1-3 Machikaneyama, Toyonaka, Osaka 560-8531, Japan}
\author{Takanori Sugiyama}
\affiliation{{\blue Research} Center for Advanced Science and Technology,
The University of Tokyo, 4-6-1 Komaba Meguro-ku, Tokyo 153-8904{\blue, Japan}}
\author{William J. Munro}
\affiliation{NTT Basic Research Laboratories, NTT Corporation, 3-1 Morinosato-Wakamiya, Atsugi, Kanagawa 243-0198, Japan}
\affiliation{NTT Theoretical Quantum Physics Center, NTT Corporation, 3-1 Morinosato-Wakamiya, Atsugi, Kanagawa 243-0198, Japan}
\affiliation{National Institute of Informatics, 2-1-2 Hitotsubashi, Chiyoda-ku, Tokyo 101-8430, Japan}

\begin{abstract}
Typically, the aim of quantum metrology is
to sense
target fields with
high precision utilizing quantum properties. Unlike the typical aim, in this paper, we use quantum properties for adding a new {\blue functionality} to quantum
sensors. More concretely, we propose a delegated quantum sensor (a client-server model) {\red with security inbuilt}. Suppose
that a client wants to measure some target fields with high precision, but he/she does not
have any high-precision sensor. This leads
the client to delegate the sensing to a remote server who possesses a high-precision sensor.
The client gives the server instructions about how to
control the sensor.
The server {\blue lets} the sensor interact with the target fields in accordance with the instructions, and then
sends the sensing measurement results to the client.
In this case, since the server knows the control process and
readout results of the sensor,
the information of the target fields is available not only for
the client but also for the server.
We show that, by using an
entanglement between the client and the server, an asymmetric
information gain is possible so that
only the client can obtain the sufficient information of the target
 fields. In our scheme, the server generates the entanglement between a solid
 state system (that can interact with the target fields) and a photon, and {\blue sends} the photon to
 the client.
On the other hand, the client {\blue is
required} to possess linear
optics elements only including wave plates, polarizing beam splitters,
 and single-photon detectors.
Our scheme is feasible with the current technology,
 and {\blue our} results pave the way for a novel application of
quantum metrology.
\end{abstract}
\maketitle

\section{Introduction}
Quantum properties such as superposition and entanglement are considered
to be useful resources for several information processing
tasks~{\red\cite{shor1997pw,grover1997quantum,harrow2009quantum,vandersypen2001experimental,bennett1984quantum,bennett1992experimental,gisin2002quantum,broadbent2009universal,morimae2013blind,
takeuchi2016blind,barz2012demonstration,greganti2016demonstration,dowling2003quantum,spiller2005introduction}}.
For example,
a quantum computer efficiently
solves some problems that seem {\blue to be} hard for classical computers~{\red\cite{shor1997pw,grover1997quantum,harrow2009quantum,vandersypen2001experimental}}.
Quantum cryptography such as quantum key distribution enables two remote
parties to communicate in an information-theoretic secure
way~\cite{bennett1984quantum,bennett1992experimental,gisin2002quantum}.
Furthermore, recently, by combining these two concepts, blind
quantum computing (BQC) protocols have also been
proposed~\cite{broadbent2009universal,morimae2013blind,takeuchi2016blind,barz2012demonstration,greganti2016demonstration}.
BQC enables a client {\blue with computationally weak devices} to delegate
{\blue universal} quantum computing to a remote server who has a universal
quantum computer while the client's privacy (input, output, and algorithm) is information-theoretically {\blue protected}.

Quantum metrology is also one of such practical applications of quantum
properties~\cite{degen2016quantum,budker2007optical,balasubramanian2008nanoscaleetal,maze2008nanoscaleetal,dolde2011electric,
neumann2013high,wineland1992dj,huelga1997improvement,matsuzaki2011magnetic,chin2012quantum}.
By using the superposition property of a qubit~\cite{degen2016quantum}, we
can improve the sensitivity to measure target fields such as
magnetic fields, electric fields, and temperature~\cite{budker2007optical,balasubramanian2008nanoscaleetal,maze2008nanoscaleetal,dolde2011electric,
neumann2013high}.
When the frequency of
the qubit can be shifted by the target fields, a superposition state
of the qubit will acquire a phase shift on the non-diagonal terms
during the interaction with the target fields. Therefore, the readout of the
phase provides us with the information of the target fields. 
Further, the use of entanglement resources enhances the measurable sensitivity, and an entanglement sensor can beat
the standard quantum limit that the sensitivity of any
classical sensor is bounded by~\cite{wineland1992dj,huelga1997improvement,matsuzaki2011magnetic,chin2012quantum}.

Just as interdisciplinary approaches between quantum computing and
quantum cryptography have lead to propose BQC, interdisciplinary
approaches between quantum metrology,
quantum computing, and quantum cryptography have lead to propose
practical quantum sensing
 protocols~\cite{kessler2014quantum,dur2014improved,arrad2014increasing,herrera2015quantum,
unden2016quantum,matsuzaki2017magnetic,higgins2007entanglement,waldherr2012high,nakayama2015quantum,
matsuzaki2017projective,komar2014quantum,eldredge2018optimal,proctor2018multiparameter}.
For example, while quantum error correction~\cite{lidar2013quantum}
is a concept that has been discussed in the field of
quantum computation for the mitigation of errors during the
computation, 
it has been found that the quantum error correction is
also useful to improve the sensitivity of quantum sensors~\cite{kessler2014quantum,dur2014improved,arrad2014increasing,herrera2015quantum,
unden2016quantum,matsuzaki2017magnetic}. A phase estimation algorithm~\cite{kitaev1997}
for quantum computation
has been used to increase the dynamic range of quantum sensors
\cite{higgins2007entanglement,waldherr2012high}.
Combination of a quantum computer and a quantum sensor provides us with
a way to implement a projective measurement of energy on target systems
\cite{nakayama2015quantum,matsuzaki2017projective}.
Besides them, although a quantum {\black network is} an important concept in quantum cryptography
\cite{sasaki2011field,wang2013direct}, a network of quantum
sensors is also becoming an attractive topic
in quantum metrology \cite{komar2014quantum,eldredge2018optimal}. This is because a quantum sensing network can
enhance the estimation precision under certain conditions
\cite{proctor2018multiparameter}.
Also, there are researches
that
combine the quantum cryptography and quantum metrology \cite{giovannetti2001quantum,giovannetti2002quantum,giovannetti2002positioning,chiribella2005optimal,chiribella2007secret,huang2017cryptographic,xie2018high}.
In the setup of these researches, a few {\blue nodes} exist, and there
are noisy  channels
 between them.
 The aim of these researches is
 to share the sensing results (measured at {\blue a node}) between the
 nodes without {\blue leaking the information to an eavesdropper} that has an access to
 the {\blue channels}.

\begin{figure}[t]
\includegraphics[width=8.5cm, clip]{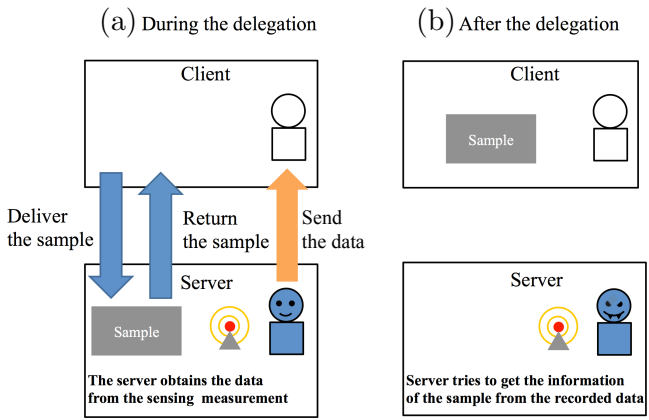}
\caption{
 A schematic diagram to illustrate {\black client-server based quantum sensing}.
 We consider the cases during and after the delegation of the sensing.
(a) During the delegation, the client sends a sample to the server who holds a high-precision quantum
sensor.
The client gives the server instructions about how to control the
quantum sensor for the measurement of the sample, and the server obeys
the instructions. The server sends the measurement results to the
client, and the server returns the sample to the client.
(b) After the delegation, the
server can try to estimate the information of the sample from the
classical data remaining in the server's quantum sensor. Interestingly, our
protocol described in this paper prevents the server from obtaining the
information of the sample from the remaining {\redd data} while the client
can recover the information of the sample, which we call
asymmetric information gain.
 }
\label{application}
\end{figure}

In this paper, we also take an interdisciplinary approach to add {\black
novel functionality} to quantum sensors.
Our protocol is based on an idea to combine the BQC and quantum metrology.
More concretely, we discuss
a situation that a client who does not have any quantum sensor tries to have an access of a remote quantum sensor that
belongs to a server at a remote site. This situation is given as Fig.~\ref{application}.
First, the client sends a sample to be measured to the server, and then gives the server instructions about how to
use the sensor for the measurement of the sample. Second, according to the instructions, the server {\blue lets} the sensor interact
with the target field (the sample). Finally, the server sends the sensing measurement results and returns the sample
to the client.
We assume here that the server obeys the client's
instructions during the delegation of quantum sensing.
 This assumption
 seems to be reasonable {\blue as} if the sample is a macroscopic
 object, it is hard to encrypt the sample.
 {\blue In fact}, if we allow the server's deviation from the instructions
 during the delegation, the server can easily obtain the information
 about the sample without being noticed by
 the client.
This is a large difference between the situation of our paper and that of BQC, since the input of BQC is typically an input of a mathematical problem.
Even under this assumption, since the classical information about the control process and
readout results of the sensor remains at the server's place after the delegation,
the information of the target field should be available not only for
the client but also for the server as shown in Fig.~\ref{application}
(b). This is problematic when the client does not want to reveal the
information of the sample.

Then, we will consider the following question: even if the classical information about
the control process and readout results of the sensor remain in the server's quantum sensor after the delegation, can we construct a protocol such that
the client obtains the sufficient information of the target field while the
server cannot do it?
We will show that such an asymmetric information gain
between the client and server
from the quantum sensor is
possible by using entanglement and a reasonably realistic
experimental setup.

{\red Standard} quantum
teleportation~\cite{bennett1993teleporting,bouwmeester1997experimental,furusawa1998unconditional}
can be a way to realize the asymmetric information gain, which, however, may have a technical
problem
in terms of feasibility as we will discuss. A superposition state of a qubit
for the sensing can acquire a relative phase due to the target
fields, and this state can be teleported to the site of the client by the
quantum teleportation if an ideal Bell pair is available between the
client's site and the server's site. In this
case, only the client knows the value of the qubit phase where
the information of target fields is encoded.
However, in this method, the client needs to keep one half of the Bell pair until the outcome of the Bell measurement is sent from the server. This means that the client should
have a quantum memory~\cite{julsgaard2004experimental}.
Although a long-lived quantum memory is possible in the state of
art technology~\cite{hedges2010efficient}, it would be more feasible if the asymmetric information gain can
be realized with less demanding conditions. Furthermore, since it
is difficult to share the ideal Bell pairs, the client has to certify
how well the server prepares the
Bell pair.
Therefore, in this paper, we will propose a more feasible scheme that can estimate the fidelity between an actual state and the ideal Bell pair, and does not
require any quantum memory for the client's site.

\begin{figure}[t]
\includegraphics[width=9cm, clip]{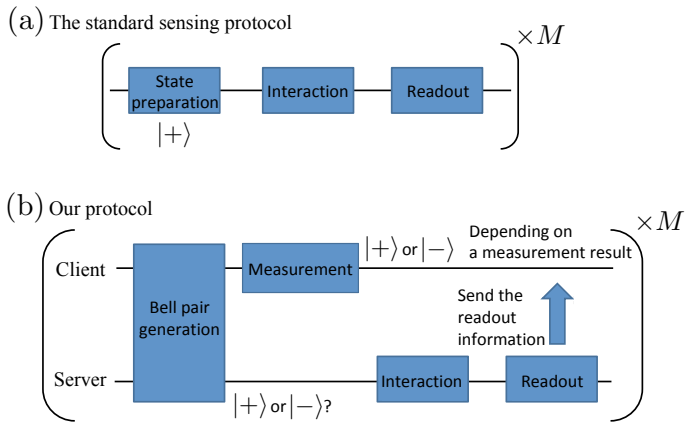}
\caption{
Comparison of the standard sensing protocol with our remote sensing
protocol. Here,
 $|\pm\rangle\equiv(|0\rangle\pm|1\rangle)/\sqrt{2}$ and $M$ is the
 repetition number.
 (a) In the standard protocol, there are three steps such as the state preparation, the interaction,
and the readout. (b)
In our protocol, the state preparation is composed of the Bell pair generation (between
the server and the client) and
a subsequent measurement by the client, while
the interaction and the readout are implemented at the server's side similar
to the standard protocol.
Since the server does not know the client's measurement result, the server cannot know {\blue which}
single-qubit state is prepared at the server's side before the
 interaction while the client can know it. This difference realizes the
 asymmetric information gain.
}
\label{circuit}
\end{figure}

We explain the basic ideas of our scheme.
As shown in Fig.~\ref{circuit}, the standard quantum metrology consists of three steps such as the state
preparation (the preparation of $|+\rangle\equiv(|0\rangle+|1\rangle)/\sqrt{2}$), the interaction with target fields, and the readout of the
qubit (for details, see Sec.~\ref{II}). In order to propose our remote sensing protocol, we divide the state
preparation step into two steps, i.e. the Bell pair generation and the client's subsequent measurement.
First, in the Bell pair generation, the client estimates the fidelity
between the actual two-qubit state $\rho$ prepared by the server and the
ideal Bell pair. If $\rho$ is close to
the ideal Bell pair, the client accepts it. Otherwise, the client rejects. Thanks to this step, we can remove the necessity of the ideal Bell pair from our protocol.
Note that in general, errors in a channel between the client and the
server vary depending on the time. Therefore, we cannot use the quantum
state tomography~\cite{smithey1993measurement,hradil1997quantum,banaszek1999maximum} and
the process tomography~\cite{poyatos1997complete,chuang1997prescription} for our purpose. 
After the client accepts $\rho$, by measuring one half of $\rho$, the
client prepares a single-qubit state at the server's side. Finally, the
server performs the standard quantum
metrology using the single-qubit state instead of $|+\rangle$, and sends the measurement result to the client.
The key idea of our protocol is that the state of the {\blue single qubit} 
is known for the client because the client knows the measurement outcome
while the server cannot know
the state.
This difference results in the asymmetric information gain where the client
obtain the sufficient information of the sample while the server does not obtain it.
Although some readers may think that the server can also obtain
the sufficient information by performing the standard quantum metrology
with initial state
$|+\rangle$ in parallel, this deviation is prohibited by our assumption mentioned above.

From an experimental point of view, our protocol would
corresponds to a situation that, after the server generates a Bell pair
between a {\blue solid-state} qubit and a flying qubit such as a photon, the
photon is sent to the client and is measured by the client immediately after the
photon arrives at the client's side. Therefore, the client does not
need any quantum memory, and the only
requirement for the client is an ability to measure the flying qubits
(photons),
which can be implemented by just basic linear
optics elements such as a wave plate, a {\blue polarizing} beam splitter, and single-photon
detectors.
{\redd In general, imperfections of projective measurements on the
qubits decrease the sensitivity as a quantum sensor. While there are
commercially available single photon detectors, an accurate projective measurement
on the solid-state qubit is not a mature technology yet, and not every researcher
is capable of implementing precise projective measurements on the solid state qubits. 
In our quantum remote sensing protocol, even when the client does not have high-precision projective measurement apparatuses of the solid-state
qubits, the client can delegate
the standard sensing protocol to the server with a technology of accurate projective measurements, and so the client can
measure the sample with better sensitivity.
To
achieve this goal, we use the quantum property such as entanglement.
}

The rest of this paper is organized as follows:
In Secs.~\ref{II} and
\ref{III}, as preliminaries, we review the standard quantum metrology
and a random-sampling test, which is a fidelity estimation protocol {\blue for}
Bell pairs. In Sec.~\ref{IV}, by combining the standard quantum
metrology and the random-sampling test, we propose {\blue the} quantum remote
sensing protocol. In
Sec.~\ref{IVA}, we give a procedure of our protocol. In Sec.~\ref{IVB},
we derive the upper bound of the uncertainty obtained by the client. In
Sec.~\ref{IVC},
we derive the lower bound of the uncertainty obtained by the
server. {\blue In Sec.~\ref{IVD}, from} these bounds of uncertainties, we show that our protocol
achieves the asymmetric
information gain. In Sec.~\ref{V}, we discuss possible experimental
implementations of our protocol. In Sec.~\ref{VI}, we conclude our discussion.

\section{Quantum metrology}
\label{II}
\begin{figure}[t]
\includegraphics[width=8cm, clip]{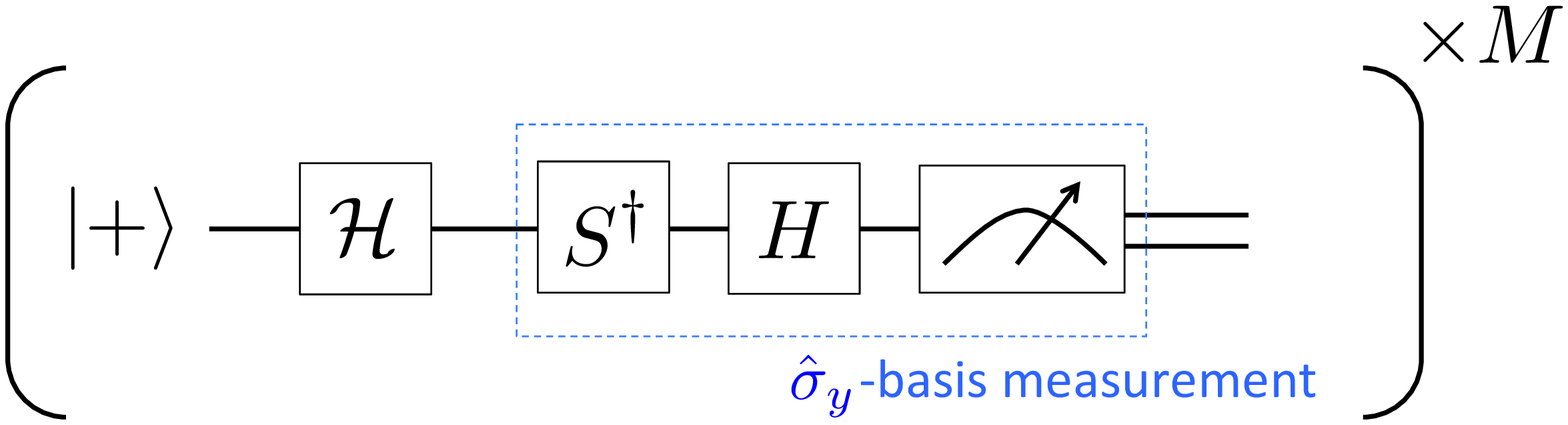}
\caption{The quantum circuit representation of the standard quantum
metrology protocol. Here, $\mathcal{H}$ represents the time evolution by the
Hamiltonian in Eq.~(\ref{hamiltonian}),
$S^\dag\equiv|0\rangle\langle0|-i|1\rangle\langle 1|$,
$H$ is the Hadamard gate, and the meter symbol represents the $\hat{\sigma}_z$-basis measurement.}
\label{smetrology}
\end{figure}

Let us review the standard quantum metrology by using a single-qubit state.
A Hamiltonian of the qubit is given as
\begin{eqnarray}
\label{hamiltonian}
\hat{\mathcal{H}}=\frac{{\black\hbar}\omega}{2}\hat{\sigma}_z,
\end{eqnarray}
where $\omega$ denotes the {\black angular} frequency of the qubit.
Suppose that we can linearly shift the resonant frequency by external
fields such as $\omega\propto B$, where $B$ denotes the amplitude of the
target field. 
By estimating the resonant frequency, we
can determine the amplitude of the target field. For such an
estimation, a typical {\blue Ramsey-type} measurement can be used. The procedure is as follows:
\begin{enumerate}
\item Prepare an initial state $|+\rangle$. 
\item Let the state $|+\rangle$
evolve by the Hamiltonian in Eq.~(\ref{hamiltonian}) for a time $t$. 
\item Measure the state by a
projection operator of
$\hat{\mathcal{P}}=(\openone+\hat{\sigma }_y)/2$.
\item Repeat steps 1-3 {\blue within} a given total time $T$.
\end{enumerate}
The number of repetitions
is described by $M=T/(t_{\rm{p}}+t+t_{\rm{r}})$ where $t_{\rm{p}}$ ($t_{\rm{r}}$)
denotes the required time for the preparation (readout) of the state.
From these repetitions, we obtain $M$ measurement
results $\{m_1,m_2, \cdots, m_M\}${\blue , where $m_j\in\{0,1\}$ $(1\le j\le M)$}. By using the average value
{\blue$S_M=(\sum_{j=1}^{M}m_j)/M$}, we can estimate the target parameter
$\omega$.
The above procedure is shown by a quantum circuit in
{\blue Fig.~\ref{smetrology}}.

The uncertainty $\delta\omega$ of the resonant frequency can be
calculated as follows{\blue :
we} have {\blue the} probability  $P\equiv {\rm Tr}[\hat{\mathcal{P}} e^{-i\hat{\mathcal{H}}t{\black/\hbar}}|+\rangle
\langle +|e^{i\hat{\mathcal{H}}t{\black/\hbar}}]$ {\blue of obtaining $m_j=1$. From Eq.~(\ref{hamiltonian}), we} obtain {\blue $P=(1+\sin \omega t)/2 \simeq
 (1+\omega t)/2$}, where we assume
 $|\omega t|\ll 1$ because we are interested in detecting a small
 amplitude of the target field. Throughout of this paper, we assume the
 same condition.
 From the average value $S_M$,
one can define an {\blue estimated} value of
$\omega $ such as {\blue$\omega
^{(\rm{est})}_M=(2S_M-1)/t$}.
We have
\begin{eqnarray*}
 \delta ^2P&=&M\langle (S_M-P)^2\rangle \\
&\simeq& M\frac{t^2}{4}\langle (\omega _M^{(\rm{est})} -\omega )^2\rangle 
\end{eqnarray*}
where $\langle \cdot \rangle $ denotes {\blue the} statistical average and $\delta
^2P=P(1-P)$ denotes {\blue the} variance. So we obtain
\begin{eqnarray*}
 \delta \omega \simeq \frac{1}{t\sqrt{M}}
\end{eqnarray*}
where $ \delta \omega \equiv \sqrt{\langle (\omega _M^{(\rm{est})}
-\omega )^2\rangle }$ denotes the uncertainty of the estimation.

\section{Random-sampling test for Bell pairs}
\label{III}
In our quantum remote sensing protocol, it is important to share the Bell pair between the client and the server. However, there are
usually many possible error sources in a channel between the client and the server,
and they decrease the fidelity {\blue with} the Bell pair. Since such {\blue the} infidelity {\blue with}
the Bell pair will affect the performance of our protocol to obtain the
asymmetric information gain (as we will describe later),
it is important to estimate a lower bound of the fidelity between an
actual two-qubit state $\rho$ prepared by the remote server and the ideal Bell
pair $|\Phi^+\rangle\equiv(|00\rangle+|11\rangle)/\sqrt{2}$. More
precisely, in order to
derive the uncertainty of the delegated quantum sensing, we have to
guarantee that the two-qubit state $\rho$
satisfies $\langle\Phi^+|\rho|\Phi^+\rangle\ge 1-\epsilon$ with
probability at least
$1-\delta$ $(0<\epsilon,\delta\le 1)$.
Since errors in the channel may vary
depending on the time, we cannot assume any independent and identically
distributed (i.i.d.) property.
This means that
the quantum
state tomography~\cite{smithey1993measurement,hradil1997quantum,banaszek1999maximum} and
the process tomography~\cite{poyatos1997complete,chuang1997prescription}
 are not appropriate
 for our purpose.
In order to estimate the fidelity without assuming any i.i.d. property,
we use a {\black destructive} random-sampling
test~\cite{NC01b,takeuchi2018resource}.
For the completeness of the paper, we review the random-sampling test for the Bell pair.

The test runs as follows:
\begin{enumerate}
\item A client sets three parameters $\epsilon$, $\delta$, and
      $\Delta$. Here, $\Delta$ determines the error-robustness of the
      random-sampling test, and $0\le\Delta<\epsilon/3$. The client
      tells these three values to the remote
      server.

\item The server sends {\black an} $8k$-qubit state $\rho_S$ to the client, where
\begin{eqnarray}
\label{cost}
k=\left\lceil\cfrac{75}{8(\epsilon-3\Delta)^2}\log{\cfrac{2}{\delta}}\right\rceil
\end{eqnarray}
with $\lceil\cdot\rceil$ {\black being} the ceiling function.
Without loss of generality, we can assume that the state $\rho_S$
      consists of $4k$ registers, and each register stores two
      qubits. If the state $\rho_S$ is not disturbed by any channel
      noise,
      $\rho_S=(|\Phi^+\rangle\langle\Phi^+|)^{\otimes 4k}$. Otherwise,
      $\rho_S$ is arbitrary $8k$-qubit quantum state whose registers may
      be entangled. 

\item The client chooses $k$ registers from $4k$ registers independently
      and uniformly {\blue at} random, and then the client measures each of them in
      the $\hat{\sigma}_x\otimes\hat{\sigma}_x$ basis, which we call the
      $X$ test. If two outcomes
      on two qubits in the same register are the same, we say the register passes the $X$ test. Otherwise, it fails the $X$ test.

\item The client chooses $k$ registers from the remaining $3k$ registers
      independently and uniformly {\blue at} random, and then the client measures
      each of them in the $\hat{\sigma}_z\otimes \hat{\sigma_z}$ basis,
      which we call the $Z$ test.
      If two outcomes on two qubits in the same register are the same, we say the register passes the $Z$ test. Otherwise, it fails the $Z$ test.

{\blue \item The client chooses one register, which we call the target register, from the remaining $2k$ registers uniformly at random. Other remaining registers are discarded.

\item The client counts the number $N_{\rm fail}$ of registers that fail the $X$ test
      or the $Z$ test. If
      $N_{\rm fail}\le 2k\Delta$, the client keeps the target register.
      Otherwise, the client discards it.}
\end{enumerate}
Note that in the random-sampling test, we do not assume any i.i.d. property of the $4k$
registers. Therefore, this test works for any channel noise. This
test also works for a certain error in the server's apparatus if it can
be treated as a channel noise.
This is why we do not have to assume that all of the server's operations are perfect.

In step {\blue 5}, $2k-1$ registers are discarded. Although this may seem
to be
a huge waste, this discarding is necessary to show
Theorem~\ref{soundness}, which is given later.
Fidelity estimation protocols that do not discard any register have
already been proposed~\cite{hayashi2015verifiable, markham2018simple},
but they have no error tolerance,
i.e. $\Delta=0$. An effective use of discarded registers has also been discussed in Ref.~\cite{takeuchi2018resource}.

In order to show that the random-sampling test works correctly as a
fidelity estimation protocol, we show two properties, so called the
completeness and the soundness. Intuitively, if the client can accept
the ideal Bell pair $|\Phi^+\rangle$ with high probability, we say that
the random-sampling test has the completeness. Thanks to the
completeness, the client does not mistakenly reject $|\Phi^+\rangle$. On
the other hand, if the random-sampling test guarantees that an accepted
quantum state is close to $|\Phi^+\rangle$ with high probability, we say
that it has the soundness. Thanks to the soundness, the client does not
mistakenly accept a quantum state that is far from
$|\Phi^+\rangle$. More rigorously, the following two theorems hold:
\begin{theorem}[Completeness]
\label{completness}
When $\rho_S=(|\Phi^+\rangle\langle\Phi^+|)^{\otimes 4k}$, the client {\blue does not discard the target register in} step 6 (i.e. the random-sampling test succeeds) with unit probability.
\end{theorem}
{\it Proof.}
The Bell pair $|\Phi^+\rangle$ is stabilized by $\hat{\sigma}_x\otimes\hat{\sigma}_x$ and $\hat{\sigma}_z\otimes\hat{\sigma}_z$. In other words, the Bell pair satisfies
\begin{eqnarray*}
\hat{\sigma}_x\otimes\hat{\sigma}_x|\Phi^+\rangle=\hat{\sigma}_z\otimes\hat{\sigma}_z|\Phi^+\rangle=|\Phi^+\rangle.
\end{eqnarray*}
Accordingly, $|\Phi^+\rangle$ always passes the $X$ test and the $Z$
test. Since the number {\blue$N_{\rm fail}$} of registers that fail the $X$ test or the $Z$
test is $0$, the client {\blue does not discard the target register in} step 6 with unit probability.
(Remember that the client {\blue does not discard the target register in} step 6 when {\blue$N_{\rm fail}\le 2k\Delta$,} and $\Delta\ge 0$.) \hspace{\fill}$\blacksquare$

\begin{theorem}[Soundness]
\label{soundness}
{\blue In step 5}, the state $\rho_{\rm tgt}$ of the target register, which is a single register chosen in step {\blue 5}, satisfies
\begin{eqnarray}
\label{Fbound}
\langle\Phi^+|\rho_{\rm tgt}|\Phi^+\rangle\ge 1-\epsilon{\blue +3\Delta-\cfrac{3N_{\rm fail}}{2k}}
\end{eqnarray}
with probability at least $1-\delta$.
\end{theorem}
{\blue In general, the client cannot determine the value of $N_{\rm fail}$ without performing experiment.
However, once $N_{\rm fail}\le 2k\Delta$ holds, Eq.~(\ref{Fbound}) gives the client the non-trivial lower bound $1-\epsilon$.}
A proof of Theorem~\ref{soundness} is given in Appendix A.

\begin{figure}[t]
\includegraphics[width=8cm, clip]{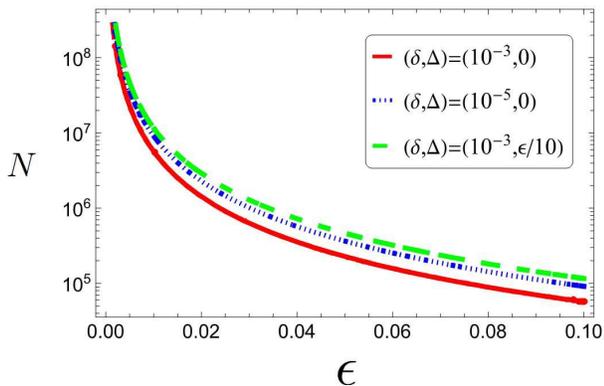}
\caption{The {\black required} number \textcolor{black}{$N=8k$} of qubits as a function of
 $\epsilon$. {\black Here, $k=\left\lceil75\log{(2/\delta)}/[8(\epsilon-3\Delta)^2]\right\rceil$.} The {\black bottom, middle, and top} lines show the value of $8k$ for
 $(\delta,\Delta)=(10^{-3},0)$, $(10^{-5},0)$,
 and $(10^{-3},\epsilon/10)$, respectively.}
\label{resource}
\end{figure}

Using Eq.~(\ref{cost}) and Theorem~\ref{soundness}, we can derive how
 many qubits are necessary to prepare a two-qubit state $\rho_{\rm tgt}$
 whose fidelity is at least $1-\epsilon$ with probability
 $1-\delta$. {\blue Note that for simplicity, we here consider the situation where $N_{\rm fail}\le 2k\Delta$ holds with unit probability.}
{\redd As examples, we explain two cases where this condition is satisfied.
First, the ideal (noiseless)
 channel can satisfy the condition. Second, we can consider a channel noise where the
 identity operation, the bit-flip operation $(\hat{\sigma}_x)$, the
 phase-flip operation $(\hat{\sigma}_z)$, or the bit- and phase-flip
 operation $(\hat{\sigma}_x\hat{\sigma}_z)$ is periodically applied.
 Suppose that Bob sends one half of a single register (two-qubit state) per a
 certain time $\tau _{\rm{B}}$, and one of three error operations
 ($\hat{\sigma}_x$, $\hat{\sigma}_z$, and
 $\hat{\sigma}_x\hat{\sigma}_z$) is applied once every
 $\tau _{\rm{B}}/\Delta$. In this case, $N_{\rm fail}$ is definitely less than or equal to $2k\Delta$.
 Note again that this situation is just a hypothetical one to simplify the discussion. Theorem~\ref{soundness} also holds for other situations.}
In
 Fig.~\ref{resource}, for some specific values of $\delta$ and $\Delta$, we show the $\epsilon$-dependence of the number $8k$ of qubits.

\section{Quantum remote sensing}
\label{IV}
In this section, as a main result, we propose a quantum remote sensing
protocol. {\black Simply speaking, our protocol runs as follows: first,
the client and the server try to share a
Bell pair. In this step, since the Bell pair is disturbed by channel
noises, they estimate the fidelity between the actual shared two-qubit
state $\rho$ and the ideal Bell pair
$|\Phi^+\rangle$ using the random-sampling test. Second, if $\rho$ is
sufficiently close to $|\Phi^+\rangle$, the client measures his/her half
of $\rho$ to prepare a single-qubit
state at the server's side. Finally, the server performs the standard
quantum metrology protocol using the single-qubit state, and then sends
the readout result to the client.
Since the server cannot know {\blue which} state is prepared by the client, our protocol achieves the asymmetric information gain.}

\subsection{Protocol}
\label{IVA}
By combining the standard quantum metrology protocol given in
Sec.~\ref{II} and the random-sampling test given in Sec.~\ref{III}, we {\black now}
propose {\blue the} quantum remote sensing protocol. In order to fit the random-sampling test to
our quantum remote sensing protocol, we slightly modify the random-sampling test. In the random-sampling test given in Sec.~\ref{III},
{\black the $8k$-qubit state} $\rho_S$ is sent simultaneously in step 2. In this
case, a quantum memory
is needed for the client. In order to remove the necessity of the
quantum memory from the client, the server sends each qubit one by one
to the client, and the client randomly chooses his/her action from the
$X$ test, $Z$ test,
discarding, and the $\hat{\sigma}_x$-basis measurement on the one half
of the target register. The $\hat{\sigma}_x$-basis measurement is
necessary to prepare a single-qubit state at the server's side,
which is used to perform the standard quantum metrology
protocol. Furthermore, we partition the $X$ test and the $Z$ test, which
are performed by the client in Sec.~\ref{III},
into the client's and the server's measurements. 

Our quantum remote sensing protocol runs as follows:
\begin{enumerate}
\item The client sets three parameters $\epsilon$, $\delta$, and $\Delta$, where $0\le\Delta<\epsilon/3$. The client tells these three values to a remote server.

\item The server prepares {\black an} $8k$-qubit state $\rho_S$, where
\begin{eqnarray*}
k=\left\lceil\cfrac{75}{8(\epsilon-3\Delta)^2}\log{\cfrac{2}{\delta}}\right\rceil
\end{eqnarray*}
{\black with} $\lceil\cdot\rceil$ {\black being} the ceiling function.
The state $\rho_S$ consists of $4k$ registers, and each registers store two qubits.

\item The client chooses $k$ registers, which we call the $X$ set, from
      $4k$ registers independently and uniformly {\blue at} random. Then, the
      client tells the server which registers are
      selected as the $X$ set.

\item The client chooses $k$ registers, which we call the $Z$ set, from
      the remaining $3k$ registers independently and uniformly {\blue at}
      random. Then, the client tells the server
      which registers are selected as the $Z$ set.

\item The client chooses one register, which we call the target
      register, from the remaining $2k$ registers independently and
      uniformly {\blue at} random. Then, the client tells the server which register
      is selected as
      the target register.

\item The server sends one half of each register to the client one by one. (In total, the server sends $4k$ qubits to the client.)

\item The client and the server perform one of following four steps for each register:
\begin{enumerate}
\item If a register is in the $X$ set, the client and the server measure
      it in the $\hat{\sigma}_x^{(C)}\otimes\hat{\sigma}_x^{(S)}$ basis,
      where the superscripts $C$ and $S$ represent an operator applied
      on the client's site
      and the server's site, respectively. If the client's and the
      server's outcomes are the same, we say the register passes the $X$
      test. Otherwise, it fails the $X$ test. Note that they can check
      whether or not their outcomes
      are the same by classical communication.

\item If a register is in the $Z$ set, the client and the server measure
      it in the $\hat{\sigma}_z^{(C)}\otimes\hat{\sigma}_z^{(S)}$
      basis. If the client's and the server's outcomes are the same, we
      say the register passes the
      $Z$ test. Otherwise, it fails the $Z$ test.

\item If a register is the target register, the client measures the one
      half of the target register in the $\hat{\sigma}_x^{(C)}$ basis to
      prepare a single-qubit state $\rho_{\rm QRS}$ at the server's
      site. Let $s\in\{0,1\}$
      be the measurement outcome. Then, the server stores $\rho_{\rm QRS}$ in his/her quantum memory.

\item Otherwise, the client and the server discard the register.
\end{enumerate}

\item The client counts the number {\blue $N_{\rm fail}$} of registers that fail the $X$ test
      or the $Z$ test. If
      {\blue $N_{\rm fail}\le 2k\Delta$}, the random-sampling test succeeds, and the client
      proceeds to the next step. Otherwise,
      the test fails, and the client aborts {\blue the protocol}.

\item
    The server performs the standard quantum metrology given in
      Sec.~\ref{II} where the initial state is replaced with the quantum state $\rho_{\rm QRS}$.
More specifically,
     the single-qubit state $\rho_{\rm QRS}$ is evolved by the
     Hamiltonian in Eq.~(\ref{hamiltonian}) and then measured in the
     $\hat{\sigma}_y$ basis.
     The server sends the measurement outcome $o\in\{0,1\}$ to the client.

\item The client calculates $s\oplus o$ and accepts it as the result of the sensing.

\item The client and the server repeat steps 1-10 $M$ times to obtain sufficiently high precision.
\end{enumerate}

We will show that our quantum remote sensing protocol achieves the asymmetric
information gain.
In other words, the uncertainty of the estimation of the
client becomes much smaller than that of the server.
Since the small uncertainty implies the success of the metrology, this
asymmetric property means that the client can obtain
more accurate information of the sensing results than
the server.
This asymmetry
comes
from the fact that the measurement outcome $s$ in step 7 (c)
is known only for the client.
To quantify such the asymmetric information gain between
the client and server,
we calculate the averaged uncertainty over the outcome $s$ from each point of view.
We assume here that each repetition is independent from the
others, which means that the probability 
of the measurement at {\blue step 9 in} a repetition has no correlation with
that {\blue in} the other repetitions.
This assumption is needed to calculate the uncertainty.
Also, due to this, we can decrease the uncertainty of
the estimation
by increasing the repetition number $M$. On the other hand,
{\blue in} the steps from 1 to 10 inside a single repetition{\blue ,}
we do not assume any i.i.d. property about the quantum
states. These assumptions seem to be reasonable especially when the
state preparation time $t_{\rm{p}}$ is much shorter than the interaction
time $t$.
More concretely, since the $8k$-qubit state $\rho_S$ is generated in a
short time, $8k$ qubits may be correlated each other. On the other hand,
since the interaction time is long, i.e. the time interval between each
repetition is long, the $8k$-qubit state $\rho_S$ prepared in the $i$th
$(1\le i\le M-1)$ repetition
should
not correlate with that prepared in the $(i+1)$th repetition.

 \begin{figure}[t]
\includegraphics[width=9cm, clip]{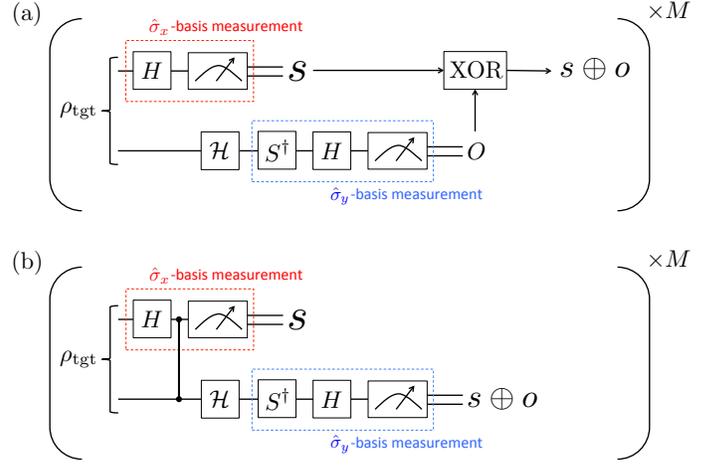}
\caption{(a) The quantum circuit representation of our delegated
 metrology where $\rho_{\rm tgt}$ denotes the state of the target register
 between the server and the client. It is worth mentioning that, by taking a large
 limit of the number $k$, $\rho_{\rm tgt}$ approaches the
 ideal Bell pair. (b) The quantum circuit equivalent to the circuit in
  (a). The vertical line represents the controlled-$\hat{\sigma}_z$
  gate {\blue $|0\rangle\langle 0|\otimes I+|1\rangle\langle 1|\otimes\hat{\sigma}_z$}.
  These two circuits output $s\oplus o$ with the same probability.}
\label{metrology}
\end{figure}

\medskip
\subsection{\red Uncertainty of the client}
\label{IVB}
{\black Let us first} calculate the averaged uncertainty of the client.
{\blue Hereafter, for simplicity, we only consider the situation where $N_{\rm fail}\le 2k\Delta$ holds with unit probability.}
To this end,
we derive the initial state $\rho_{\rm QRS}$ of the standard quantum
metrology protocol. We give a quantum circuit corresponding to
 our quantum remote sensing protocol in Fig.~\ref{metrology} (a). By
 transforming the quantum circuit in Fig.~\ref{metrology} (a), we obtain
 the quantum circuit in Fig.~\ref{metrology} (b). Since these two
 quantum circuits
 output $s\oplus o$ with the same probability, the uncertainties
 calculated from these two quantum circuits are also the same. Let $p_s$
 and $\rho^{(s)}$ be the probability of the client obtaining the
 measurement outcome $s$
 and the single-qubit state prepared when the measurement outcome is
 $s$, respectively. In Fig.~\ref{metrology} (b),
 $\sum_{s=0}^1p_s\hat{\sigma}_z^s\rho^{(s)}\hat{\sigma}_z^s$ is used as
 the initial state of the standard
 quantum metrology protocol. Therefore, we can assume that $\rho_{\rm
 QRS}=\sum_{s=0}^1p_s\hat{\sigma}_z^s\rho^{(s)}\hat{\sigma}_z^s$. Note
 that the quantum circuit in Fig.~\ref{metrology} (b) does not achieve
 the asymmetric
 information gain because the value of $s\oplus o$ is revealed also for
 the server. However, our present aim is to calculate the
 uncertainty. Therefore, we can replace the quantum circuit in
 Fig.~\ref{metrology}
 (a) with that in Fig.~\ref{metrology} (b).

 \begin{figure}[t]
\includegraphics[width=8cm, clip]{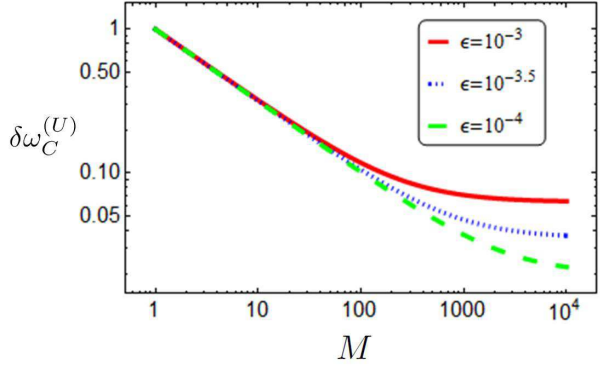}
\caption{The client's uncertainty $\delta \omega _{{C}}^{({U})}$against
 the number $M$ of repetitions where we set $t=1$. Although the horizontal axis represents
 discrete values, we use continuous lines for {\blue plots} as guides for the eyes.
}
\label{newuncertaintyclientafter}
\end{figure}

From Theorem~\ref{soundness}, the target register satisfies
$\langle\Phi^+|\rho_{\rm tgt}|\Phi^+\rangle\ge 1-\epsilon$ with
probability at least $1-\delta$. Therefore, for the virtual initial
state
$\sum_{s=0,1}p_s\hat{\sigma}_z^s\rho^{(s)}\hat{\sigma}_z^s$, the following corollary holds:
\begin{corollary}
\label{coro}
$\langle+|\sum_{s=0,1}p_s\hat{\sigma}_z^s\rho^{(s)}\hat{\sigma}_z^s|+\rangle\ge
 1-\epsilon$ with a probability at least $1-\delta$.
\end{corollary}
{\it Proof.} We use the monotonicity of the fidelity, i.e. a property
that the fidelity is not decreased by any trace-preserving (TP)
map. First, we consider the non-demolition $\hat{\sigma}_x$-basis measurements on the first qubits of $|\Phi^+\rangle$ and $\rho_{\rm
tgt}$. Note that these non-demolition measurements are not performed in
practice. These are only used to show this corollary.
Then, the Hadamard gates are applied on the first qubits. As a result, $|\Phi^+\rangle$ and $\rho_{\rm tgt}$ becomes
\begin{eqnarray}
\label{state}
\cfrac{|0+\rangle\langle 0+|+|1-\rangle\langle 1-|}{2}
\end{eqnarray}
and 
\begin{eqnarray}
\label{state2}
p_0|0\rangle\langle 0|\otimes \rho^{(0)}+p_1|1\rangle\langle 1|\otimes \rho^{(1)},
\end{eqnarray}
respectively. Here, $|-\rangle\equiv(|0\rangle-|1\rangle)/\sqrt{2}$. Next, we perform the controlled-$\hat{\sigma}_z$ gates, which
correspond to $\hat{\sigma}_z^s$ applied on the second qubits, and then
trace out the first qubits. After these operations, Eqs.~(\ref{state})
and (\ref{state2})
become $|+\rangle$ and
\begin{eqnarray*}
\sum_{s=0,1}p_s\hat{\sigma}_z^s\rho^{(s)}\hat{\sigma}_z^s,
\end{eqnarray*}
respectively. Since the above four operations (the non-demolition
measurement, the Hadamard gate, the controlled-$\hat{\sigma}_z$ gate,
and the discarding of the first qubit) are
TP maps, from Theorem~\ref{soundness},
\begin{eqnarray*}
\langle
 +|\sum_{s=0,1}p_s\hat{\sigma}_z^s\rho^{(s)}\hat{\sigma}_z^s|+\rangle\ge\langle\Phi^+|\rho_{\rm
 tgt}|\Phi^+\rangle\ge 1-\epsilon
\end{eqnarray*}
with a probability at least $1-\delta$.
\hspace{\fill}$\blacksquare$

From Corollary~\ref{coro}, we derive the upper bound of the averaged
uncertainty $\delta\omega_C$ obtained by the client as follows:
\begin{theorem}
 \label{uncertaintyc}
{\blue Let $\epsilon<1/2$. Then, in} the limit of small $\omega$,
\begin{eqnarray*}
 \delta \omega _C
 \leq \frac{1}{t} {\blue\sqrt{\frac{1}{M}+4(\epsilon -\epsilon ^2)}} \equiv \delta\omega_C^{(U)} \label{clientbound}
\end{eqnarray*}
with a probability at least $(1-\delta)^M$.
 \end{theorem}
A proof of {\blue Theorem~\ref{uncertaintyc}} is given in Appendix B.

Although the client should
have a probability  $P= {\rm Tr}[\hat{\mathcal{P}} e^{-i\hat{\mathcal{H}}t{\black/\hbar}}|+\rangle
\langle +|e^{i\hat{\mathcal{H}}t{\black/\hbar}}]$ with {\blue the ideal Bell
pair}, the actual probability $P'={\rm
Tr}[\hat{\mathcal{P}} e^{-i\hat{\mathcal{H}}t{\black/\hbar}}
(\sum_{s=0,1}p_s\hat{\sigma}_z^s\rho^{(s)}\hat{\sigma}_z^s)e^{i\hat{\mathcal{H}}t{\black/\hbar}}]$ is not known for the
client due to the possible errors in the channel between the client and
server. This lack of the knowledge of the exact form of the probability 
induces a residual error that is not reduced by increasing
the number $M$ of the repetitions.
We plot the uncertainty
 $\delta\omega_C^{(U)}$ against $M$ in
 Fig.~\ref{newuncertaintyclientafter}, and this actually shows that the
 uncertainty is bounded by the residual error even {\blue when $M$ is large}.
 As we decrease $\epsilon
 $, $\delta\omega_C^{(U)}$ approaches to $\delta \omega $.

\subsection{Uncertainty of the server}
\label{IVC}
Next, we calculate the
uncertainty of the server.
The server cannot know the value of $s$. Therefore, from the viewpoint
of the server, $\rho_{\rm\black QRS}={\rm Tr}_C[\rho_{\rm{tgt}}]$, where ${\rm
Tr}_C[\cdot]$ is the partial trace over the qubit possessed by the
client.
 Since ${\rm Tr}_C[|\Phi^+\rangle\langle\Phi^+|]=\openone/2$, from Theorem~\ref{soundness}, the fidelity
 between the completely mixed state $\openone/2$ and $\rho_{\rm\black QRS}$ is at least $1-\epsilon$ with probability at least $1-\delta$.

{\blue From this fact}, we show the following theorem:
\begin{theorem}
 \label{uncertaintys}
Let $\delta\omega_S$ be the uncertainty of the server. Then, in the limit of small $\omega$,
\begin{eqnarray*}
 \delta \omega _S\geq \frac{1}{2t}\sqrt{\frac{1-4(\epsilon
   -\epsilon ^2)}{M(\epsilon -\epsilon ^2)}} \equiv  \delta \omega _S^{(L)}
\end{eqnarray*}
 with a probability at least $(1-\delta
 )^M$.
 \end{theorem}
A proof of Theorem~\ref{uncertaintys} is given in Appendix C.

Here, {\blue in order to decrease the server's uncertainty $\delta\omega_S^{(L)}$ as much as possible,} we assume that the server knows the actual
form of the probability  $P'$ unlike the client. In this case, the
uncertainty {\blue$\delta\omega_S^{(L)}$} of the estimation decreases as we increase the repetition
number $M$.

\subsection{Comparison of the {\blue uncertainties} between the client and the
  server}
\label{IVD}

\begin{figure}[t]
\includegraphics[width=8cm, clip]{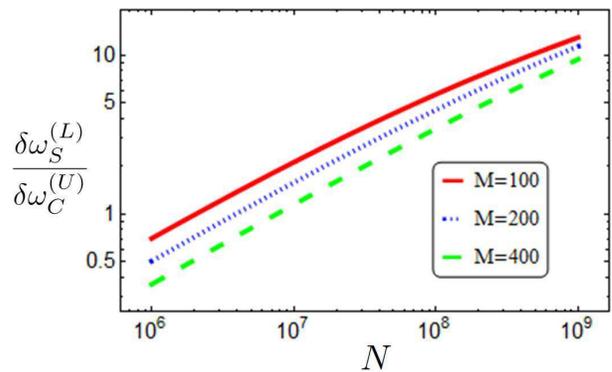}
\caption{
   Plot of $\delta\omega_S^{(L)}/\delta\omega_C^{(U)}$ against
the number $N=8k$ of {\blue qubits required to extract a single two-qubit state that is close to the} Bell pair
 by the random-sampling test where we set $\delta =10^{-6}$ and $\Delta
 =0$.  Although the horizontal axis represents
 discrete values, we use continuous lines for {\blue plots} as guides for the eyes.
 }
\label{newnewasymmetryst}
\end{figure}

\begin{figure}[t]
 \includegraphics[width=8cm, clip]{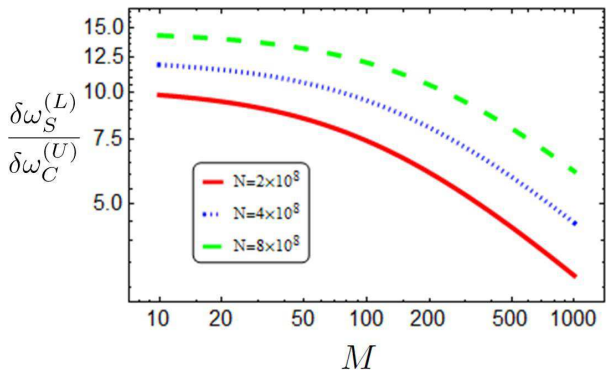}
 \caption{
  Plot of $\delta\omega_S^{(L)}/\delta\omega_C^{(U)}$ against
the number $M$ of repetitions
  where we set $\delta
  =10^{-6}$ and $\Delta =0$. Although the horizontal axis represents
  discrete values, we use continuous lines for {\blue plots} as guides for the eyes.
  }
 \label{newnewnewasymmetryst}
 \end{figure}

Using Theorems~\ref{uncertaintyc} and \ref{uncertaintys},
We compare the
uncertainty of the client and that of the server.
In our protocol, we use
$8k=8\left\lceil75\log{(2/\delta)}/[8(\epsilon-3\Delta)^2]\right\rceil$
qubits to generate the {\blue single-qubit} state $\rho_{\rm QRS}$. This
means that we have $\epsilon \simeq 3\Delta
+\sqrt{75\log{(2/\delta)}/8k}$. From this relationship, we
can plot the ratio $\delta\omega_S^{(L)}/\delta\omega_C^{(U)}$ against
$N=8k$ (the number of the qubits to {\blue extract} a single high-fidelity Bell
pair) as shown in Fig.~\ref{newnewasymmetryst}.
Importantly,
as we increase $N$ {\blue that corresponds to decrease $\epsilon$}, we can increase the ratio
$\delta\omega_S^{(L)}/\delta\omega_C^{(U)}$, and so the information gain
becomes more asymmetric.
Also, we plot the ratio
$\delta\omega_S^{(L)}/\delta\omega_C^{(U)}$ against $M$ in
Fig.~\ref{newnewnewasymmetryst}. As we increase the number $M$ of repetitions, the uncertainty of the client becomes closer to that of
the server. This comes from {\blue the fact} that the server does not
know the precise {\blue form of the} probability $P'$. {\blue However}, Fig.~\ref{newnewnewasymmetryst} shows that, by taking $M\leq 1000$, we can
realize a large asymmetric information gain such as
$\delta\omega_S^{(L)}/\delta\omega_C^{(U)}\geq 5$ {\blue when $\Delta=0$, $\delta=10^{-6}$, and $8k=8\times 10^8$}.

{\redd From Fig.~\ref{newuncertaintyclientafter}, by increasing the value of $M$, we can decrease the client's uncertainty $\delta\omega_C$.
However, from Fig.~\ref{newnewnewasymmetryst}, we notice that the
asymmetric information gain becomes smaller as we increase $M$. In fact,
$\delta\omega_S^{(L)}/\delta\omega_C^{(U)}$
is a monotonically decreasing function of $M$, which is known from
Theorems~\ref{uncertaintyc} and \ref{uncertaintys}. Furthermore, the
probability $(1-\delta)^M$ of
Theorems~\ref{uncertaintyc} and \ref{uncertaintys} is exponentially decreased as we increase $M$.
Therefore, in order to achive the sufficiently large
$\delta\omega_S^{(L)}/\delta\omega_C^{(U)}$, the sufficiently large
$(1-\delta)^M$, and the sufficiently small $\delta\omega_C$
simultaneously, we have to make the value of $N$ sufficiently large. In
other words, by increasing $N$, we can achieve the arbitrary large
asymmetric information gain even when $M$
is quite large. We can observe this behaviour in Fig.~\ref{newnewnewasymmetryst}.}

{\blue In the quantum metrology, it is common to use the standard
deviation as the measure of the uncertainty. However, it is not the only
way to compare the uncertainties between
the client and the server. In Appendix D, to investigate the asymmetric
information gain more deeply, we discuss another method to evaluate the
uncertainties and obtain
the similar asymmetric information gain.}

\section{Possible experimental realization}
\label{V}
We discuss possible experimental realizations of our protocol.
To implement our protocol, we need a solid state system that has a strong
coupling with the target fields. Also, a quantum transducer from the
solid state system to photons is required to generate a Bell
pair between the client and the server. There are
several systems that satisfy these requirements.

Nitrogen vacancy (NV) center in diamond is one of the candidates to
realize our protocol~\cite{maze2008nanoscaleetal, taylor2008high, balasubramanian2008nanoscaleetal,Go01a,gruber1997scanning,jelezko2002single}.
NV centers provide us with a spin triplet, and we can use this system as
an effective two level system by using frequency
selectivity. Microwave pulses allow us to implement single-qubit gate
operations of the NV centers~\cite{JGPGW01a}.
We can readout the state of the NV centers through photoluminescence detection~\cite{jiang2009repetitive}.
On top of these properties, the NV centers have a coupling with magnetic
fields, and so the
NV centers can be used to measure
magnetic fields with a high sensitivity~\cite{maze2008nanoscaleetal,
taylor2008high, balasubramanian2008nanoscaleetal}.
Moreover, the NV centers are considered as a candidate to realize a
distributed quantum computer and a quantum repeater
{\red\cite{Barrett:2005p363,childress2006fault,nemoto2014photonic,nemoto2016photonic}}.
Actually, an entanglement between the NV center and a flying
photon can be generated with the current technology~\cite{togan2010quantum}.
These properties are prerequisite for the possible realization of our
protocol.

Moreover, there is a practical motivation to use our
scheme of a 
delegated quantum sensor with the NV center. It is preferable to
fabricate a smaller sensor to improve the spatial resolution.
Although many
efforts have been made to create the NV center in a small nanodiamond
\cite{rondin2012nanoscale,cuche2009near,balasubramanian2008nanoscaleetal,maletinsky2012robust},
the fabrication of the small nanodiamond containing an NV center
is not a mature technology yet. So the client who wants to measure the
sample with a high spatial resolution can delegate the sensing to the
server who is capable of fabricating such a nanodimaond with a NV center.

A superconducting flux qubit (FQ) coupled with electron spins would be
another candidate to realize our protocol~\cite{twamley2010superconducting,marcos2010coupling,zhu2011coherent,zhudark2014,matsuzaki2015improving}.
High fidelity gate
operations of the FQ are available with the current technology~\cite{bylander2011noise}, and it is possible to
implement quantum non-demolition measurements on the FQ~\cite{ClarkeWilhelm01a}.
Moreover, the FQ can be a sensitive {\blue magnetic-field} sensor due to the
large persistent current of the FQ~\cite{bal2012ultrasensitive}. It is worth mentioning that the FQ
itself does not have a direct coupling with the optical photons. However, the quantum
state of the FQ can be transferred to the electron spins~\cite{saito2013towards}, and some of the electron spins such as NV
centers or rare-earth doped crystals have a coupling with the
photons. By using these properties, it is in principle possible to
convert the quantum information encoded in the FQ into the form of the
photons~\cite{blum2015interfacing,lai2018single,o2014interfacing}.
Although such a quantum transducer from the superconducting qubits to the
photons is not experimentally demonstrated yet, such a hybrid approach
can also be a candidate to realize our remote sensing protocol in the
future.

Also, there would be practical advantage for the client
to use our delegation scheme with the FQ. The resonant frequency $\omega _{\rm{FQ}}$ of the
FQ can be shifted by the applied magnetic field $B$, and the derivative has
a linear relationship with the persistent current $I_{\rm{P}}$ of the FQ such as
$d\omega _{\rm{FQ}}/dB\propto I_{\rm{P}}$. So the FQ with a higher persistent current has a
better sensitivity as a magnetic field sensor. However, such a realization of the high persistent
current requires a special design of the superconducting circuit
\cite{paauw2009tuning,twamley2010superconducting}, and not every
researcher could fabricate such a sample. In our delegation
scheme, if the server has an ability to fabricate such a high persistent
current FQ, the client can use the server's FQ to improve the
sensitivity to measure the sample that the client has.

\section{Conclusion}
\label{VI}
We have proposed a delegated quantum sensing protocol.
We have considered a situation where the client asks the
server to measure the target sample when the client does not have a quantum
sensor but the server does. The client provides a sample to be measured
with the server, and {\blue sends} the server
the instructions about how
to use the quantum sensor to measure the sample.
\textcolor{black}{The server obeys the client's
instructions during the measurement of the sample, and will return the
sample to the client after the measurement.}
Importantly, \textcolor{black}{if
the standard quantum sensing scheme is naively implemented,}
not only the client but
also the server can obtain the information of the client's sample \textcolor{black}{even
after returning the sample, because of the server's knowledge of the
measurement results.}
We show that, by
using an entanglement between the client and the server, it is possible to
realize an asymmetric information gain where
only the client can obtain the sufficient information of the sample while the
server cannot do it. 
Our protocol does not require any
quantum memory for the client, and so our protocol would be feasible
even in the current technology.

However, the resource cost of our protocol is higher than that of
the standard quantum sensing protocol, because we need approximately
$10^9$ qubits in our parameter regime as shown in Fig.~\ref{newnewnewasymmetryst}.
As a future work, it is
interesting to propose a more
resource-efficient quantum remote sensing protocol.

\section*{ACKNOWLEDGMENTS}
We thank Koji Azuma and Shiro Saito for helpful discussions.
This work was supported
by CREST (JPMJCR1774), JST and Program for Leading Graduate Schools: Interactive Materials Science Cadet Program, and in part by
MEXT Grants-in-Aid for Scientific Research on Innovative Areas ``Science of
hybrid quantum systems'' (Grant No. 15H05870).
Y. T. and Y. M. contributed equally to this work.

\section*{APPENDIX A: PROOF OF THEOREM~\ref{soundness}}
In this Appendix, we give a proof of Theorem~\ref{soundness}.\\
{\it Proof.}
The proof is similar to that of Ref.~\cite{takeuchi2018resource}.
In order to show this theorem, we use the following inequality:
\begin{lemma}[Serfling's bound~\cite{serfling1974probability,tomamichel2017largely}]
\label{randomsampling}
Consider a set of binary random variables $Y=(Y_1,Y_2,\ldots,Y_T)$ with $Y_j$ $(1\le j\le T)$ taking values in $\{0,1\}$ and $T=N+K$. Then, for any $0<\nu<1$,
\begin{eqnarray*}
{\rm Pr}\left[\sum_{j\in\bar{\Pi}}Y_j\ge \cfrac{N}{K}\sum_{j\in\Pi}Y_j+N\nu\right]\le{\rm exp}\left[-\cfrac{2\nu^2NK^2}{(N+K)(K+1)}\right],
\end{eqnarray*}
where $\Pi$ is a set of $K$ samples chosen independently and uniformly at random from $Y$ without replacement. $\bar{\Pi}$ is the complementary set of $\Pi$.
\end{lemma}
Note that the sampling without replacement means that once a sample is selected, it is removed from the population in all subsequent selections.

Let $\Pi^X$ and $\Pi^Z$ be the sets of $k$ registers used for the $X$
test and the $Z$ test, respectively. In step 3 of the random-sampling
test,
the client measures $k$ registers of the set $\Pi^X$ in the
$\hat{\sigma}_x\otimes\hat{\sigma}_x$ basis. If the $j$th $(1\le j\le
k)$ register passes
the $X$ test, we set $Y_j=0$. Otherwise, $Y_j=1$. Therefore, by setting
$K=k$ and $T=4k$ in Lemma~\ref{randomsampling}, the number
$\sum_{j\in\bar{\Pi}^X}Y_j$ of
registers that are not stabilized by $\hat{\sigma}_x\otimes\hat{\sigma}_x$, where $\bar{\Pi}^X$ is the complementary set of $\Pi^X$, is upper bounded by
\begin{eqnarray}
\label{Xnumber}
3k\nu+3\sum_{j\in\Pi^X}Y_j
\end{eqnarray}
with probability at least
\begin{eqnarray}
\label{Xprob}
1-{\rm exp}\left[-\cfrac{6\nu^2k^3}{4k(k+1)}\right]\equiv q_X.
\end{eqnarray}

Next, in step 4 of the random-sampling test, the client measures $k$
registers of the set $\Pi^Z$ in the
$\hat{\sigma}_z\otimes\hat{\sigma}_z$ basis. Note that since the
registers used for the $Z$ test are selected from registers
that are not used for the $X$ test, $\bar{\Pi}^X\supset \Pi^Z$. By
setting $Y_j$ in the similar manner to the case of the $X$ test, and
setting $K=k$ and $T=3k$ in Lemma~\ref{randomsampling}, the number
$\sum_{j\in\bar{\Pi}^Z}Y_j$
of registers that are not stabilized by $\hat{\sigma}_z\otimes\hat{\sigma}_z$, where $\bar{\Pi}^Z$ is the set of remaining $2k$ registers, is upper bounded by
\begin{eqnarray}
\label{Znumber}
2k\nu+2\sum_{j\in\Pi^Z}Y_j
\end{eqnarray}
with probability at least
\begin{eqnarray}
\label{Zprob}
1-{\rm exp}\left[-\cfrac{4\nu^2k^3}{3k(k+1)}\right]\equiv q_Z.
\end{eqnarray}

We set $\nu=2(\epsilon-3\Delta)/5$. {\blue From} Eqs.~(\ref{Xnumber}), (\ref{Xprob}), (\ref{Znumber}), and
(\ref{Zprob}), we can guarantee that among
the remaining $2k$ registers, at least
\begin{eqnarray}
\nonumber
&&(4k-2k)-\left(3k\nu+3\sum_{j\in\Pi^X}Y_j\right)-\left(2k\nu+2\sum_{j\in\Pi^Z}Y_j\right)\\
\nonumber
&=&2k-5k\nu-3\sum_{j\in\Pi^X}Y_j-2\sum_{j\in\Pi^Z}Y_j\\
\nonumber
&\ge&2k-5k\nu-{\blue 3N_{\rm fail}}\\
\nonumber
&=&2k-2k(\epsilon-3\Delta)-{\blue 3N_{\rm fail}}\\
\label{correct}
&=&2k{\blue \left(1-\epsilon+3\Delta-\cfrac{3N_{\rm fail}}{2k}\right)}
\end{eqnarray}
registers always pass the $X$ test and the $Z$ test simultaneously with probability at least
\begin{eqnarray}
\nonumber
&&q_Xq_Z\\
\nonumber
&=&\left[1-{\rm exp}\left(-\cfrac{3}{2}\nu^2k\cfrac{1}{1+1/k}\right)\right]\\
\nonumber
&&\times\left[1-{\rm exp}\left(-\cfrac{4}{3}\nu^2k\cfrac{1}{1+1/k}\right)\right]\\
\label{qq1}
&\ge &\left[1-{\rm exp}\left(-\cfrac{3}{4}\nu^2k\right)\right]\left[1-{\rm exp}\left(-\cfrac{2}{3}\nu^2k\right)\right]\\
\nonumber
&\ge&1-2{\rm exp}\left(-\cfrac{2}{3}\nu^2k\right)\\
\label{qq2}
&\ge&1-2{\rm exp}\left(-\log{\cfrac{2}{\delta}}\right)\\
\label{probability}
&=&1-\delta,
\end{eqnarray}
where we have used $k>1$ and Eq.~(\ref{cost}) to derive Eqs.~(\ref{qq1}) and (\ref{qq2}){\blue , respectively}.
Only the ideal Bell pair $|\Phi^+\rangle$ can always passes both of the
$X$ test and the $Z$ test. Therefore, from Eq.~(\ref{correct}), the
ratio of the number of the ideal Bell pairs to that of non-ideal
two-qubit quantum states
in the remaining $2k$ registers is at least
\begin{eqnarray}
\label{correct2}
{\blue \cfrac{2k\left[1-\epsilon+3\Delta-3N_{\rm fail}/(2k)\right]}{2k}=1-\epsilon+3\Delta-\cfrac{3N_{\rm fail}}{2k}}.
\end{eqnarray}

Let 
\begin{eqnarray*}
\Bigg\{|\beta_{ij}\rangle\equiv (\hat{\sigma}_z^i\otimes \openone)(\hat{\sigma}_x^j\otimes \openone)\cfrac{|00\rangle+|11\rangle}{\sqrt{2}}\Bigg|i,j\in\{0,1\}\Bigg\}
\end{eqnarray*}
be the Bell basis.
The uniform random selection in step {\blue 5} is equivalent to selecting the
first register of the remaining $2k$ registers after the random
permutation. Therefore, Eqs.~(\ref{probability}) and (\ref{correct2})
mean that when the state
$\rho_{\rm tgt}$ is expanded by the Bell basis:
\begin{eqnarray*}
\rho_{\rm tgt}=\sum_{i,j}\sum_{i',j'}q_{iji'j'}|\beta_{ij}\rangle\langle\beta_{i'j'}|,
\end{eqnarray*}
where {\blue$\sum_{i,j}q_{ijij}=1$}, the coefficient $q_{0000}$
is at least {\blue $1-\epsilon+3\Delta-3N_{\rm fail}/(2k)$} with probability at least
$1-\delta$. Accordingly, from $|\beta_{00}\rangle=|\Phi^+\rangle$ and
$\langle\beta_{ij}|\Phi^+\rangle=0$
for any $(i,j)\neq(0,0)$,
\begin{eqnarray*}
\langle \Phi^+|\rho_{\rm tgt}|\Phi^+\rangle\ge1-\epsilon{\blue +3\Delta-\cfrac{3N_{\rm fail}}{2k}}
\end{eqnarray*}
with probability at least $1-\delta$.
\hspace{\fill}$\blacksquare$

\section*{APPENDIX B: PROOF OF THEOREM~\ref{uncertaintyc}}
In this Appendix, we give a proof of Theorem~\ref{uncertaintyc}.\\
{\it Proof.}
First, let us describe a general theory about how to derive the
uncertainty of the estimation. Suppose that a state $\rho _{\omega }${\blue , which is the $|+\rangle$ state in the standard quantum metrology,} is
given, and this state is measured by a projective measurement. We {\blue here} assume
that the probability has a linear dependence on
$\omega$. {\blue Therefore, we} have
\begin{eqnarray*}
P= {\rm{Tr}}[\hat{\mathcal{P}}\rho _{\omega }]=x +y \omega.
\end{eqnarray*}
If $M$ copies of the state $\rho _{\omega }$ are given, one can measure
the state $M$ times and obtain {\blue $M$} measurement results {\blue$\{m_j\}_{j=1}^M$, where $m_j\in\{0,1\}$}. From the average value
{\blue $S_M=(\sum_{j=1}^{M}m_j)/M$}, one can estimate the value of
$\omega $ such as {\blue $\omega^{(\rm{est})}_M=(S_M-x)/y$}.
However, if an unknown error occurs, the actual given state might be
$\rho '_{\omega }$ that is different from $\rho _{\omega }$,
\textcolor{black}{and we consider this case}. Suppose
that the probability for this state is described as 
\begin{eqnarray*}
P'= {\rm{Tr}}[\hat{\mathcal{P}}\rho '_{\omega }]=x' +y' \omega,
\end{eqnarray*}
and this is the actual probability in this case.
{\blue Furthermore, the average value becomes $S'_M=(\sum_{j=1}^Mm'_j)/M$, where $m'_j (\in\{0,1\})$ is the measurement result obtained from the $j$th copy of $\rho'_\omega$. Let ${\omega'}_M^{({\rm est})}\equiv (S'_M-x)/y$. From the difference between $P$ and $P'$, we}
have
{\blue\begin{eqnarray}
 \delta ^2P'&\equiv&M \langle (S'_M- P')^2\rangle \nonumber \\
 &=&M \langle (x+y{\omega'}_M^{(\rm{est})}-x'-y'\omega )^2\rangle
  \nonumber \\
 &=&M \left\langle \left[ \left( x-x'\right)+y ({\omega'}_M^{(\rm{est})}
               -\omega ) +(y-y')\omega 
\right]^2
  \right\rangle. \nonumber
\end{eqnarray}}
So we obtain
\begin{eqnarray}
\nonumber
&& \delta \omega _C \\
&=& \sqrt{\langle ({\blue {\omega'}_M^{(\rm{est})}}-\omega
 )^2\rangle }\nonumber \\
 &=&\sqrt{\frac{\frac{\delta ^2 P'}{M}+(x-x')^2 +(y-y')^2
                        \omega ^2 +2(x-x')(y-y')\omega }{y^2}}\nonumber
\end{eqnarray}
where we {\blue have used $\langle {\omega'}^{(\rm{est})}\rangle =(\langle S'_M\rangle
-x)/y=(x'+y'\omega 
-x)/y$}. By considering small $\omega $, we obtain
\begin{eqnarray}
 \delta \omega _{\rm{C}}\simeq \frac{1}{y}\sqrt{\frac{\delta ^2P'}{M}+(x-x')^2}. \label{comebackuncertainty}
\end{eqnarray}

Next, {\blue we adopt the above general theory to calculate the uncertainty of the client in our protocol}.
We define $\rho_0
\equiv\sum_{s=0,1}p_s\hat{\sigma}_z^s\rho^{(s)}\hat{\sigma}_z^s$, 
and we
obtain $F\equiv\langle
+|\rho _0|+\rangle \geq 1-\epsilon $ with probability at least $1-\delta$
from Corollary~\ref{coro}.
Since any single-qubit state can be written \textcolor{black}{by a sum of Pauli matrices,
we obtain}
\begin{eqnarray*}
\rho _0=\cfrac{\openone+r_x\hat{\sigma}_x+r_y\hat{\sigma}_y+r_z\hat{\sigma}_z}{2},
\end{eqnarray*}
where $r_x$, $r_y$, and $r_z$ are real values such that $r_x^2+r_y^2+r_z^2\le 1$.
We can derive the upper bound {\blue $\delta\omega_C^{(U)}$} of the client's uncertainty by optimizing
the values of $r_x$, $r_y$, and $r_z$ under the condition that
$F\geq 1-\epsilon$.
{\blue From}
\begin{eqnarray}
&& F=\langle +|\rho _0|+\rangle =\frac{1}{2}+\frac{r_x}{2}\geq 1-
 \epsilon, \nonumber 
\end{eqnarray}
we obtain $r_x\geq 1-2\epsilon $ and $r_y^2 \leq
1-r_x^2-r_z^2\leq 1- r_x^2 \leq 1-(1-2\epsilon )^2 =4\epsilon
-4\epsilon ^2$.
{\blue We here assume that $\epsilon<1/2$.}
{\blue On the other hand, from Eq.~(\ref{hamiltonian}),} the time evolution is described by
\begin{eqnarray*}
\rho (t) \simeq  \rho _0-i \frac{\omega t}{2}[\hat{\sigma }_z,\rho
 _0]
\end{eqnarray*}
where we consider a small $\omega $. {\blue Therefore, we} have
\begin{eqnarray*}
 {\rm {Tr}}[\rho (t) \hat{\sigma }_y]\simeq r_y+\omega t r_x.
\end{eqnarray*}
So
we obtain
\begin{eqnarray}
\label{Pprime}
 P'={\rm {Tr}}\left[\frac{\openone+\hat{\sigma }_y}{2}\rho (t)\right]\simeq 
  \frac{1+r_y+\omega t r_x}{2}.
\end{eqnarray}
{\blue Since $\delta^2P'=P'(1-P')$, from Eqs.~(\ref{comebackuncertainty}) and (\ref{Pprime}),}
\begin{eqnarray}
  \delta^2 \omega _C=\frac{1}{y^2}\left [\frac{\delta
   ^2P'}{M}+(x-x')^2\right ]\leq \frac{1}{t^2} \left
   [\frac{1}{M}+4(\epsilon -\epsilon ^2) \right ], \nonumber
\end{eqnarray}
where {\blue we have used $x=1/2$, $y=t/2$, $x'=(1+r_y)/2$, and $|r_y|\leq 2\sqrt{\epsilon
-\epsilon ^2}$.}
So we obtain
\begin{eqnarray*}
   \delta \omega ^{(U)}_C= \frac{1}{t}
   \sqrt{\frac{1}{M}+4\left(\epsilon -\epsilon ^2\right)}
\end{eqnarray*}
as the upper bound {\blue on} the estimation error for the client.
The above argument is true if $\langle +|\rho_0|+\rangle\ge 1-\epsilon$ holds for all $M$ repetitions. Since this holds with probability at least $1-\delta$ for each repetition, {\blue and each repetition is assumed to be independent from the other repetiions,} the above argument is true with probability at least {\blue $(1-\delta)^M$}.
\hspace{\fill}$\blacksquare$

\section*{APPENDIX C: PROOF OF THEOREM~\ref{uncertaintys}}
In this Appendix, we give a proof of Theorem~\ref{uncertaintys}.\\
{\it Proof.}
We can describe the single-qubit state $\rho_{\rm\black QRS}$ of the server as follows:
\begin{eqnarray*}
 \rho_{\rm\black QRS}=\cfrac{\openone+R_x\hat{\sigma}_x+R_y\hat{\sigma}_y+R_z\hat{\sigma}_z}{2}.
\end{eqnarray*}
{\blue Let $F(\openone/2,\rho_{\rm QRS})$ be the
 fidelity
 between $\openone/2$ and $\rho_{\rm QRS}$.}
We derive the lower bound {\blue $\delta\omega_S^{(L)}$} of the server's uncertainty by optimizing the values of $R_x$, $R_y$, and $R_z$ under the condition that $F(\openone/2,\rho_{\rm\black QRS})\ge1-\epsilon$. 
From Eq.~(\ref{hamiltonian}), we have
\begin{eqnarray}
P&=&{\rm{Tr}}\left[e^{-i\hat{\mathcal{H}}t{\black/\hbar}}\rho_{\rm\black
QRS}e^{i\hat{\mathcal{H}}t{\black/\hbar}}\frac{\openone+\hat{\sigma
}_y}{2}\right]\nonumber
\\
&=&\frac{1+R_y\cos \omega t +R_x\sin \omega
t}{2}\nonumber 
\\
 &\simeq &\frac{1+R_y+R_x \omega
t}{2}
\label{serverprobability}
\end{eqnarray}
for a small $\omega $.
In {\blue the} polar coordinate, we consider $R_x=R \sin \theta \cos \phi $,
$R_y=R \sin \theta \sin \phi $, and $R_z=R \cos \theta $, where $R\equiv\sqrt{R_x^2+R_y^2+R_z^2}$, $0\le\theta\le\pi$, and $0\le\phi< 2\pi$,
On the other hand, we obtain
\begin{eqnarray*}
F(\openone/2,\rho_{\rm QRS})=\frac{1}{2}+\frac{1}{2}\sqrt{1-R^2}\geq
1-\epsilon 
\end{eqnarray*}
and so we have $2\sqrt{\epsilon -\epsilon ^2}\geq R\geq R_x$ {\blue and $2\sqrt{\epsilon -\epsilon ^2}\geq R_y$}.
We assume that the server knows the form of the density
matrix {\blue $\rho_{\rm QRS}$}, and so the server can estimate the value of $\omega $ from the
actual {\blue form of the probability $P$} unlike the client.
{\blue From Eqs.~(\ref{comebackuncertainty}) and (\ref{serverprobability}),}
\begin{eqnarray*}
  \delta^2 \omega _S\simeq \frac{1}{y^2}\frac{\delta
   ^2P'}{M}\simeq\frac{1-R_y^2}{t^2R_x^2M}\geq \frac{1-4(\epsilon
   -\epsilon ^2)}{4t^2M(\epsilon -\epsilon ^2)},
\end{eqnarray*}
where {\blue we have used
$y=tR_x/2$, $R_x\leq  2\sqrt{\epsilon -\epsilon ^2} $, and $R_y\leq  2\sqrt{\epsilon -\epsilon ^2} $.}
So we obtain
\begin{eqnarray*}
\delta \omega _S^{(L)}=\frac{1}{2t}\sqrt{\frac{1-4(\epsilon
   -\epsilon ^2)}{M(\epsilon -\epsilon ^2)}}
\end{eqnarray*}
as the lower bound {\blue on} the uncertainty for the server.
The above argument is true if $F(\openone/2,\rho_{\rm QRS})\ge
1-\epsilon$ holds for all $M$ repetitions.
Since this holds with probability at least $1-\delta$ for each repetition, {\blue and each repetition is assumed to be indepedent from the other repetitions,} the above argument is true with probability at least {\blue $(1-\delta)^M$}.
\hspace{\fill}$\blacksquare$

\section*{APPENDIX D: ANOTHER WAY TO EVALUATE THE UNCERTAINTY OF THE ESTIMATION}
In the quantum metrology, it is common to
use the variance (or the standard deviation) for the estimation of the
accuracy of the sensing protocol. This is the reason why we have adopted
this measure to evaluate the asymmetric information gain between the
client and server about the sensing results in the main text.
However, {\blue in some other communities, the
variance is not the standard measure but people typically use a statistical
inequality}.

In this Appendix, we evaluate the uncertainty of the estimation using
{\blue Hoeffding's inequality (for details, see Lemma~\ref{Hoeffding})}. Although such {\blue the}
estimation is not common in the field of the quantum metrology, we
include these results especially for the readers who are {\blue familiar
with statistics}.

\subsection{The standard quantum metrology}
First, let us consider the simple protocol to measure the target field
using a single qubit. The setup is the same as that described in
Sec.~\ref{II}.
{\red We evaluate a confidence interval for estimaion results of quantum metrology as analyzed in Ref.~\cite{sugiyama2015precision}.
We use Hoeffding's inequality for simplicity, although an empirical Hoeffding inequality was used in Ref.~\cite{sugiyama2015precision}.
We also limit the parameter region into a range that the linearization explained in Appendix B is valid.
The uncertainty in the approach is defined as}
\begin{eqnarray*}
 {\red\Delta} \omega \equiv | \omega ^{(\rm{est})}_M-\omega
 |\nonumber
 =\frac{2}{t}|S_M-P|.
\end{eqnarray*}
To evaluate this uncertainty, we will use the following inequality:
\begin{lemma}[Hoeffding's inequality \cite{hoeffding1963probability}]
\label{Hoeffding}
Consider a set of {\blue independent} random variables $X=(X_1,X_2,\ldots,X_M)$ with $X_j$ $(1\le j\le M)$ taking values in {\blue the interval $[0,1]$}. Then, for any $\mu\ge 0$,
{\blue\begin{eqnarray*}
{\rm Pr}\left[\left|\cfrac{1}{M}\sum_{j=1}^M X_j-\left\langle \cfrac{1}{M}\sum_{j=1}^M X_j \right\rangle\right| \geq \mu
        \right]\le 2e^{-2\mu ^2M}
\end{eqnarray*}}
where $\langle \cdot \rangle$ denotes the expectation value.
\end{lemma}
{\blue From Lemma~\ref{Hoeffding} with $\mu=\tilde{s}/\sqrt{M}$, for any positive $\tilde{s}$ satisfying $e^{-2{\tilde{s}}^2}\le 1/2$},
\begin{eqnarray*}
 {\red\Delta} \omega
 \leq \frac{2\tilde{s}}{t\sqrt{M}} \label{standardsensing}
\end{eqnarray*}
{\blue holds} with a probability at least $1-2e^{-2\tilde{s}^2}$.

\subsection{Uncertainty of the client}
We will explain how to derive the uncertainty of the client using Hoeffding's inequality.
The setup is the same as that described in Sec. \ref{IVB}.
First, let us consider a general case.
While one estimates the value of $\omega $ based on the
ideal probability
\begin{eqnarray*}
P= {\rm{Tr}}[\hat{\mathcal{P}}\rho _{\omega }]=x +y \omega ,
\end{eqnarray*}
the actual probability (that may be deviated from the ideal one due to unknown errors) is
described as
\begin{eqnarray*}
P'= {\rm{Tr}}[\hat{\mathcal{P}}\rho '_{\omega }]=x' +y' \omega .
\end{eqnarray*}
We can calculate the uncertainty as follows:
{\blue\begin{eqnarray}
\nonumber
&& {\red\Delta} \omega \\
&\equiv& | {\omega'} ^{(\rm{est})}_M-\omega
 |\nonumber \\
 &=&\left|\frac{S'_M-x}{y}-\frac{P'-x'}{y'}\right|\nonumber \\
 &= &\left|\frac{y'S'_M -yP' -xy'+x'y}{yy'}\right|\nonumber \\
  &= &\left|\frac{y'S'_M-yS'_M +yS'_M -yP' +x'y-xy +xy-xy'}{yy'}\right|\nonumber \\
   &\leq &\frac{|y'-y||S'_M| +|y||S'_M -P'|+|y||x'-x|
                 +|x||y-y'|}{|yy'|}  \nonumber
   \\ \label{newformula}
\end{eqnarray}}
Next, we {\blue adopt the above general theory to calculate the uncertainty} {\red$\Delta\omega_C$} of the client.
We have {\blue $x=1/2$, $y=t/2$, $x'=(1+r_y)/2$,
$y'=tr_x/2$, $r_x\geq 1-2\epsilon $, $|r_y|\leq 2\sqrt{\epsilon
-\epsilon ^2}$, $|y-y'|\leq |t\epsilon| $, and $|x-x'|\leq
\sqrt{\epsilon -\epsilon ^2}$}.
By substituting these into Eq{\red .}~(\ref{newformula}), we obtain
\begin{eqnarray}
\nonumber
 &&{\red\Delta} \omega _C \\
&=&|{\omega'}_M^{\rm{(est)}}-\omega
 |\nonumber \\
 &\leq &\frac{|y'-y||S'_M| +|y||S'_M -P'|+|y||x'-x|
                 +|x||y-y'|}{|yy'|}\nonumber \\
 &\leq &\frac{2}{t(1-2\epsilon )}(2\epsilon
  |S'
_M|+|S'_M-P'|+ \sqrt{\epsilon -\epsilon
  ^2}+\epsilon )\nonumber \\
 &\leq &\frac{2}{t(1-2\epsilon )}(3\epsilon
  +|S'_M-P'|+ \sqrt{\epsilon -\epsilon
  ^2} ).
\label{clientbound} 
\end{eqnarray}
{\blue The above argument is true with probability at least $(1-\delta)^M$ from the same reason as Appendix B.}

We will use the inequality described in Lemma~\ref{Hoeffding}
with Eq.~(\ref{clientbound}), and we obtain{\blue , for any positive $\tilde{s}$ satisfying $e^{-2{\tilde{s}}^2}\le 1/2$,}
\begin{eqnarray*}
 {\red \Delta} \omega _{\red C}
 \leq \frac{2}{t(1-2\epsilon )}\left(3\epsilon
  +\frac{\tilde{s}}{\sqrt{M}}+ \sqrt{\epsilon -\epsilon
  ^2} \right){\blue \equiv{\red\Delta}\omega_C^{(U)}}
\end{eqnarray*}
with a probability at least {\blue $(1-2e^{-2\tilde{s}^2})(1-\delta)^M$}, where we
choose $\mu =\tilde{s}/\sqrt{M}$.

As we explained above, the actual probability {\blue $P'$} is not known for the
client, and this lack of the knowledge 
induces a residual error that is not reduced by increasing
the number $M$ of the repetitions.

\subsection{Uncertainty of the server}
 \begin{figure}[t]
\includegraphics[width=8cm, clip]{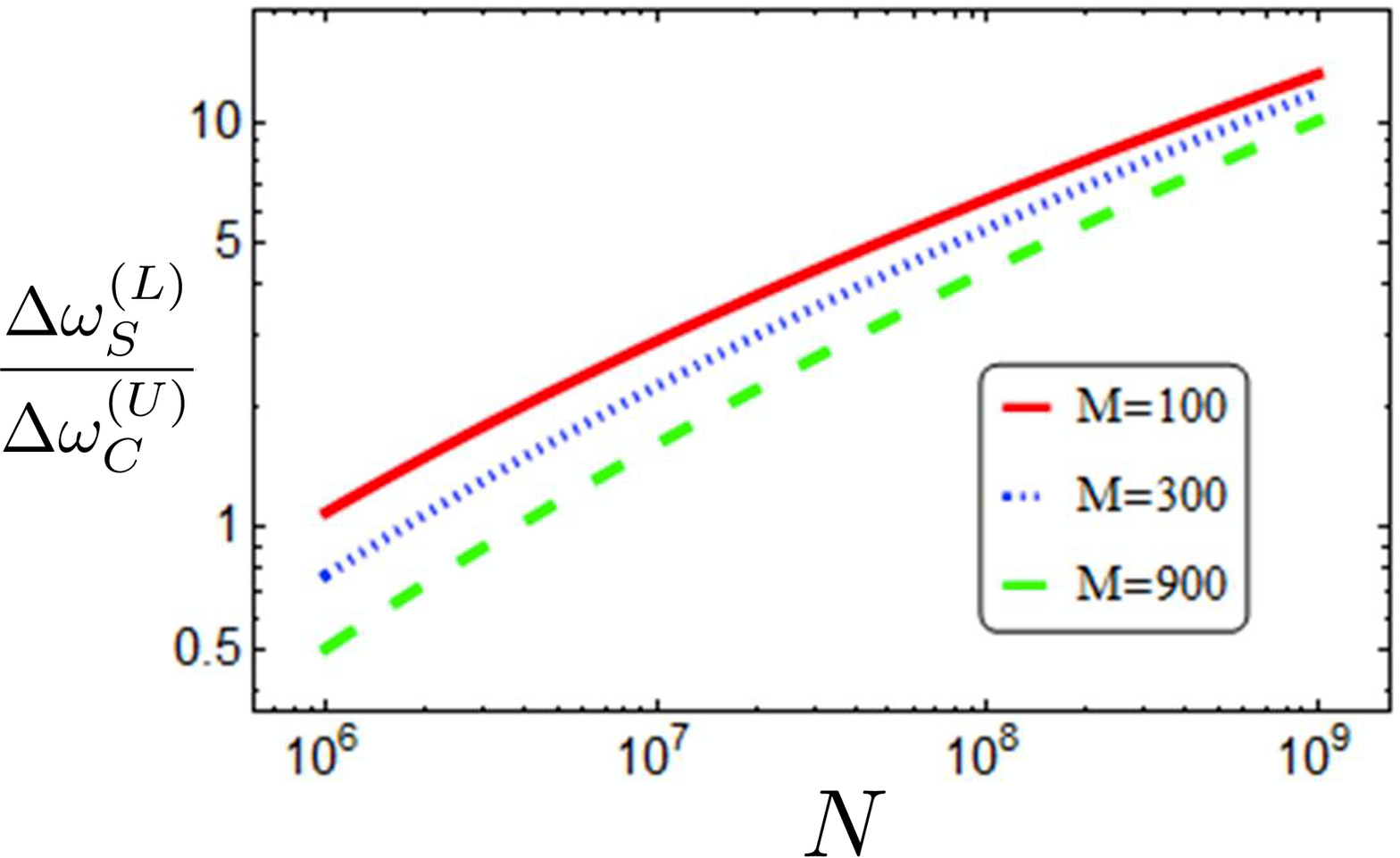}
\caption{Plot of ${\red\Delta}\omega_S^{(L)}/{\red\Delta}\omega_C^{(U)}$ against
 the number $N=8k$ of {\blue qubits required to extract a single two-qubit state that is close to the} Bell pair
 by the random-sampling test where we set $\delta =10^{-6}$, $\Delta
 =0$, and $\tilde{s}=2$ Although the horizontal axis represents
 discrete values, we use continuous lines for {\blue plots} as guides for the eyes.
 }
\label{newasymmetryst}
\end{figure}

From Eqs.~(\ref{serverprobability}) and (\ref{newformula}), we obtain
${\red\Delta} \omega _S\equiv |\omega ^{\rm{(est)}}_M-\omega|
 = 2|S_M -P|/(R_xt)$.
Since we consider the worst case where the server obtains the
largest amount of the information, it is natural to choose the parameter $R_x$
that minimizes the uncertainty {\red $\Delta\omega_S$ of the server}. {\blue As a result, we} have
\begin{eqnarray}
 \min {\red\Delta} \omega _S= \frac{|S_M -P|}{t\sqrt{\epsilon -\epsilon ^2}}\label{minimumerrorserverts}
\end{eqnarray}
where we {\blue have used} $R_x\leq 2\sqrt{\epsilon -\epsilon ^2}$.
{\blue This argument is tru with probability at least $(1-\delta)^M$ from the same reason as Appendix C.}

\begin{figure}[t]
\includegraphics[width=8cm, clip]{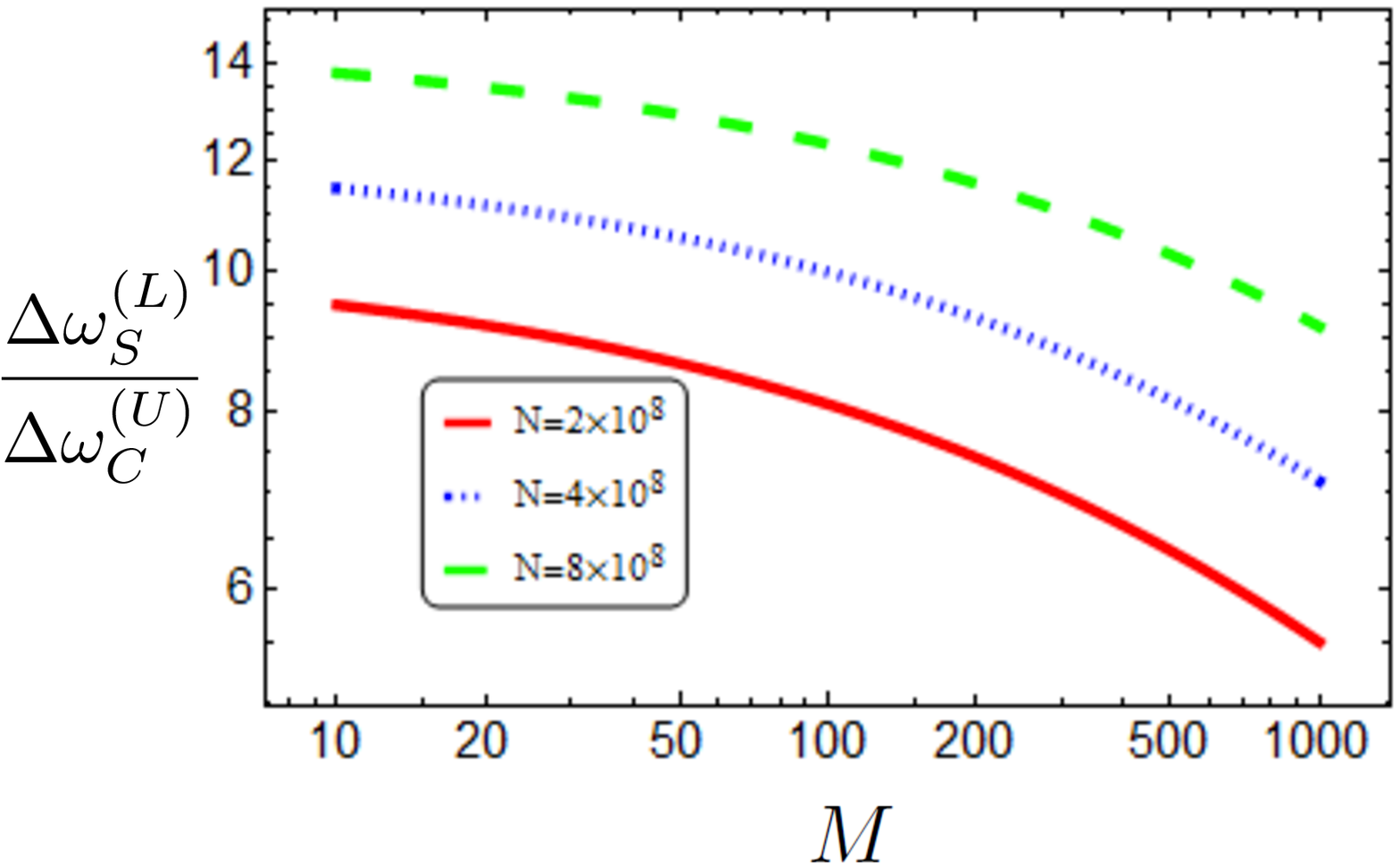}
\caption{Plot of ${\red\Delta}\omega_S^{(L)}/{\red\Delta}\omega_C^{(U)}$ against
 \textcolor{black}{the number $M$ of repetitions
 where we set $\delta
 =10^{-6}$, $\Delta =0$, and \textcolor{black}{$\tilde{s}=2$}. Although the horizontal axis represents
 discrete values, we use continuous lines for {\blue plots} as guides for the eyes.}
 }
\label{asymmetryst}
\end{figure}

When the server tries to estimate the value
of $\omega $ with the optimal state, the server obtains the uncertainty described in
Eq.~(\ref{minimumerrorserverts}). Then, the server needs to choose {\blue a statistical}
inequality to evaluate $|S_M -P|$. Although there are many choices,
we consider a case of Hoeffing's inequality. 
By using Lemma~\ref{Hoeffding}, we obtain{\blue , for any positive $\tilde{s}$ satisfying $e^{-2\tilde{s}^2}\leq 1/2$,}
\begin{eqnarray}
\min {\red\Delta} \omega _S
\leq  \frac{\tilde{s}}{t\sqrt{M}\sqrt{\epsilon -\epsilon ^2}}{\blue \equiv{\red\Delta}\omega_S^{(L)}}
\label{servererrorsthoeffding}
\end{eqnarray}
with a probability {\blue $(1-2e^{-2\tilde{s}^2})(1-\delta)^M$}.
It is worth mentioning that, if the server could use another inequality,
the server might achieve a better value than that described in
Eq.~(\ref{servererrorsthoeffding}).
However, finding the best inequality for the server is beyond the scope
of this paper, and so we leave this point as a future work.

\subsection{Comparison between the client and server}
We compare the
uncertainty of the client and that of the server.
We plot the ratio ${\red\Delta}\omega_S^{(L)}/{\red\Delta}\omega_C^{(U)}$ against
$N=8k$ in Fig.~\ref{newasymmetryst}.
As we increase $N$, we can increase the ratio
${\red\Delta}\omega_S^{(L)}/{\red\Delta}\omega_C^{(U)}$, and so the information gain
becomes more asymmetric.
Also, we plot the ratio
${\red\Delta}\omega_S^{(L)}/{\red\Delta}\omega_C^{(U)}$ against $M$ in
Fig.~\ref{asymmetryst}. 
As we increase the number $M$ of repetitions, the uncertainty of the client becomes closer to that of
the server, because the {\blue client} does not
know the precise {\blue form of the} probability $P'$. 
These results qualitatively agree with the results shown in
Figs.~\ref{newnewasymmetryst} and \ref{newnewnewasymmetryst} where we use the
standard deviation for the evaluation.

\bibliography{7mylibrary}

\begin{thebibliography}{93}
\expandafter\ifx\csname natexlab\endcsname\relax\def\natexlab#1{#1}\fi
\expandafter\ifx\csname bibnamefont\endcsname\relax
  \def\bibnamefont#1{#1}\fi
\expandafter\ifx\csname bibfnamefont\endcsname\relax
  \def\bibfnamefont#1{#1}\fi
\expandafter\ifx\csname citenamefont\endcsname\relax
  \def\citenamefont#1{#1}\fi
\expandafter\ifx\csname url\endcsname\relax
  \def\url#1{\texttt{#1}}\fi
\expandafter\ifx\csname urlprefix\endcsname\relax\def\urlprefix{URL }\fi
\providecommand{\bibinfo}[2]{#2}
\providecommand{\eprint}[2][]{\url{#2}}

\bibitem[{\citenamefont{Shor}(1997)}]{shor1997pw}
\bibinfo{author}{\bibfnamefont{P.~W.} \bibnamefont{Shor}},
  \bibinfo{journal}{SIAM J. Comput.} \textbf{\bibinfo{volume}{26}},
  \bibinfo{pages}{1484} (\bibinfo{year}{1997}).

\bibitem[{\citenamefont{Grover}(1997)}]{grover1997quantum}
\bibinfo{author}{\bibfnamefont{L.~K.} \bibnamefont{Grover}},
  \bibinfo{journal}{Phys. Rev. Lett.} \textbf{\bibinfo{volume}{79}},
  \bibinfo{pages}{325} (\bibinfo{year}{1997}).

\bibitem[{\citenamefont{Harrow et~al.}(2009)\citenamefont{Harrow, Hassidim, and
  Lloyd}}]{harrow2009quantum}
\bibinfo{author}{\bibfnamefont{A.~W.} \bibnamefont{Harrow}},
  \bibinfo{author}{\bibfnamefont{A.}~\bibnamefont{Hassidim}}, \bibnamefont{and}
  \bibinfo{author}{\bibfnamefont{S.}~\bibnamefont{Lloyd}},
  \bibinfo{journal}{Phys. Rev. Lett.} \textbf{\bibinfo{volume}{103}},
  \bibinfo{pages}{150502} (\bibinfo{year}{2009}).

\bibitem[{\citenamefont{Vandersypen et~al.}(2001)\citenamefont{Vandersypen,
  Steffen, Breyta, Yannoni, Sherwood, and
  Chuang}}]{vandersypen2001experimental}
\bibinfo{author}{\bibfnamefont{L.~M.} \bibnamefont{Vandersypen}},
  \bibinfo{author}{\bibfnamefont{M.}~\bibnamefont{Steffen}},
  \bibinfo{author}{\bibfnamefont{G.}~\bibnamefont{Breyta}},
  \bibinfo{author}{\bibfnamefont{C.~S.} \bibnamefont{Yannoni}},
  \bibinfo{author}{\bibfnamefont{M.~H.} \bibnamefont{Sherwood}},
  \bibnamefont{and} \bibinfo{author}{\bibfnamefont{I.~L.}
  \bibnamefont{Chuang}}, \bibinfo{journal}{Nature (London)}
  \textbf{\bibinfo{volume}{414}}, \bibinfo{pages}{883} (\bibinfo{year}{2001}).

\bibitem[{\citenamefont{Bennett and Brassard}(1984)}]{bennett1984quantum}
\bibinfo{author}{\bibfnamefont{C.~H.} \bibnamefont{Bennett}} \bibnamefont{and}
  \bibinfo{author}{\bibfnamefont{G.}~\bibnamefont{Brassard}}, in
  \emph{\bibinfo{booktitle}{Proceedings of IEEE International Conference on
  Computers, Systems, and Signal Processing}} (\bibinfo{organization}{IEEE, New
  York}, \bibinfo{year}{1984}), p. \bibinfo{pages}{175}.

\bibitem[{\citenamefont{Bennett et~al.}(1992)\citenamefont{Bennett, Bessette,
  Brassard, Salvail, and Smolin}}]{bennett1992experimental}
\bibinfo{author}{\bibfnamefont{C.~H.} \bibnamefont{Bennett}},
  \bibinfo{author}{\bibfnamefont{F.}~\bibnamefont{Bessette}},
  \bibinfo{author}{\bibfnamefont{G.}~\bibnamefont{Brassard}},
  \bibinfo{author}{\bibfnamefont{L.}~\bibnamefont{Salvail}}, \bibnamefont{and}
  \bibinfo{author}{\bibfnamefont{J.}~\bibnamefont{Smolin}},
  \bibinfo{journal}{Journal of cryptology} \textbf{\bibinfo{volume}{5}},
  \bibinfo{pages}{3} (\bibinfo{year}{1992}).

\bibitem[{\citenamefont{Gisin et~al.}(2002)\citenamefont{Gisin, Ribordy,
  Tittel, and Zbinden}}]{gisin2002quantum}
\bibinfo{author}{\bibfnamefont{N.}~\bibnamefont{Gisin}},
  \bibinfo{author}{\bibfnamefont{G.}~\bibnamefont{Ribordy}},
  \bibinfo{author}{\bibfnamefont{W.}~\bibnamefont{Tittel}}, \bibnamefont{and}
  \bibinfo{author}{\bibfnamefont{H.}~\bibnamefont{Zbinden}},
  \bibinfo{journal}{Reviews of modern physics} \textbf{\bibinfo{volume}{74}},
  \bibinfo{pages}{145} (\bibinfo{year}{2002}).

\bibitem[{\citenamefont{Broadbent et~al.}(2009)\citenamefont{Broadbent,
  Fitzsimons, and Kashefi}}]{broadbent2009universal}
\bibinfo{author}{\bibfnamefont{A.}~\bibnamefont{Broadbent}},
  \bibinfo{author}{\bibfnamefont{J.}~\bibnamefont{Fitzsimons}},
  \bibnamefont{and} \bibinfo{author}{\bibfnamefont{E.}~\bibnamefont{Kashefi}},
  in \emph{\bibinfo{booktitle}{Proceedings of the 50th Annual Symposium on
  Foundations of Computer Science}} (\bibinfo{organization}{IEEE, Los
  Alamitos}, \bibinfo{year}{2009}), p. \bibinfo{pages}{517}.

\bibitem[{\citenamefont{Morimae and Fujii}(2013)}]{morimae2013blind}
\bibinfo{author}{\bibfnamefont{T.}~\bibnamefont{Morimae}} \bibnamefont{and}
  \bibinfo{author}{\bibfnamefont{K.}~\bibnamefont{Fujii}},
  \bibinfo{journal}{Phys. Rev. A} \textbf{\bibinfo{volume}{87}},
  \bibinfo{pages}{050301(R)} (\bibinfo{year}{2013}).

\bibitem[{\citenamefont{Takeuchi et~al.}(2016)\citenamefont{Takeuchi, Fujii,
  Ikuta, Yamamoto, and Imoto}}]{takeuchi2016blind}
\bibinfo{author}{\bibfnamefont{Y.}~\bibnamefont{Takeuchi}},
  \bibinfo{author}{\bibfnamefont{K.}~\bibnamefont{Fujii}},
  \bibinfo{author}{\bibfnamefont{R.}~\bibnamefont{Ikuta}},
  \bibinfo{author}{\bibfnamefont{T.}~\bibnamefont{Yamamoto}}, \bibnamefont{and}
  \bibinfo{author}{\bibfnamefont{N.}~\bibnamefont{Imoto}},
  \bibinfo{journal}{Phys. Rev. A} \textbf{\bibinfo{volume}{93}},
  \bibinfo{pages}{052307} (\bibinfo{year}{2016}).

\bibitem[{\citenamefont{Barz et~al.}(2012)\citenamefont{Barz, Kashefi,
  Broadbent, Fitzsimons, Zeilinger, and Walther}}]{barz2012demonstration}
\bibinfo{author}{\bibfnamefont{S.}~\bibnamefont{Barz}},
  \bibinfo{author}{\bibfnamefont{E.}~\bibnamefont{Kashefi}},
  \bibinfo{author}{\bibfnamefont{A.}~\bibnamefont{Broadbent}},
  \bibinfo{author}{\bibfnamefont{J.~F.} \bibnamefont{Fitzsimons}},
  \bibinfo{author}{\bibfnamefont{A.}~\bibnamefont{Zeilinger}},
  \bibnamefont{and} \bibinfo{author}{\bibfnamefont{P.}~\bibnamefont{Walther}},
  \bibinfo{journal}{Science} \textbf{\bibinfo{volume}{335}},
  \bibinfo{pages}{303} (\bibinfo{year}{2012}).

\bibitem[{\citenamefont{Greganti et~al.}(2016)\citenamefont{Greganti, Roehsner,
  Barz, Morimae, and Walther}}]{greganti2016demonstration}
\bibinfo{author}{\bibfnamefont{C.}~\bibnamefont{Greganti}},
  \bibinfo{author}{\bibfnamefont{M.-C.} \bibnamefont{Roehsner}},
  \bibinfo{author}{\bibfnamefont{S.}~\bibnamefont{Barz}},
  \bibinfo{author}{\bibfnamefont{T.}~\bibnamefont{Morimae}}, \bibnamefont{and}
  \bibinfo{author}{\bibfnamefont{P.}~\bibnamefont{Walther}},
  \bibinfo{journal}{New Journal of Physics} \textbf{\bibinfo{volume}{18}},
  \bibinfo{pages}{013020} (\bibinfo{year}{2016}).

\bibitem[{\citenamefont{Dowling and Milburn}(2003)}]{dowling2003quantum}
\bibinfo{author}{\bibfnamefont{J.~P.} \bibnamefont{Dowling}} \bibnamefont{and}
  \bibinfo{author}{\bibfnamefont{G.~J.} \bibnamefont{Milburn}},
  \bibinfo{journal}{Philos. Trans. R. Soc. London A}
  \textbf{\bibinfo{volume}{361}}, \bibinfo{pages}{1655} (\bibinfo{year}{2003}).

\bibitem[{\citenamefont{Spiller et~al.}(2005)\citenamefont{Spiller, Munro,
  Barrett, and Kok}}]{spiller2005introduction}
\bibinfo{author}{\bibfnamefont{T.~P.} \bibnamefont{Spiller}},
  \bibinfo{author}{\bibfnamefont{W.~J.} \bibnamefont{Munro}},
  \bibinfo{author}{\bibfnamefont{S.~D.} \bibnamefont{Barrett}},
  \bibnamefont{and} \bibinfo{author}{\bibfnamefont{P.}~\bibnamefont{Kok}},
  \bibinfo{journal}{Contemp. Phys.} \textbf{\bibinfo{volume}{46}},
  \bibinfo{pages}{407} (\bibinfo{year}{2005}).

\bibitem[{\citenamefont{Degen et~al.}(2016)\citenamefont{Degen, Reinhard, and
  Cappellaro}}]{degen2016quantum}
\bibinfo{author}{\bibfnamefont{C.}~\bibnamefont{Degen}},
  \bibinfo{author}{\bibfnamefont{F.}~\bibnamefont{Reinhard}}, \bibnamefont{and}
  \bibinfo{author}{\bibfnamefont{P.}~\bibnamefont{Cappellaro}},
  \bibinfo{journal}{arXiv:1611.02427}  (\bibinfo{year}{2016}).

\bibitem[{\citenamefont{Budker and Romalis}(2007)}]{budker2007optical}
\bibinfo{author}{\bibfnamefont{D.}~\bibnamefont{Budker}} \bibnamefont{and}
  \bibinfo{author}{\bibfnamefont{M.}~\bibnamefont{Romalis}},
  \bibinfo{journal}{Nat. Phys.} \textbf{\bibinfo{volume}{3}},
  \bibinfo{pages}{227} (\bibinfo{year}{2007}).

\bibitem[{\citenamefont{Balasubramanian~{\it et
  al.}}(2008)}]{balasubramanian2008nanoscaleetal}
\bibinfo{author}{\bibfnamefont{G.}~\bibnamefont{Balasubramanian~{\it et al.}}},
  \bibinfo{journal}{Nature (London)} \textbf{\bibinfo{volume}{455}},
  \bibinfo{pages}{648} (\bibinfo{year}{2008}).

\bibitem[{\citenamefont{Maze~{\it et al.}}(2008)}]{maze2008nanoscaleetal}
\bibinfo{author}{\bibfnamefont{J.}~\bibnamefont{Maze~{\it et al.}}},
  \bibinfo{journal}{Nature (London)} \textbf{\bibinfo{volume}{455}},
  \bibinfo{pages}{644} (\bibinfo{year}{2008}).

\bibitem[{\citenamefont{Dolde~{\it et al.}}(2011)}]{dolde2011electric}
\bibinfo{author}{\bibfnamefont{F.}~\bibnamefont{Dolde~{\it et al.}}},
  \bibinfo{journal}{Nat. Phys.} \textbf{\bibinfo{volume}{7}},
  \bibinfo{pages}{459} (\bibinfo{year}{2011}).

\bibitem[{\citenamefont{Neumann~{\it et al.}}(2013)}]{neumann2013high}
\bibinfo{author}{\bibfnamefont{P.}~\bibnamefont{Neumann~{\it et al.}}},
  \bibinfo{journal}{Nano letters} \textbf{\bibinfo{volume}{13}},
  \bibinfo{pages}{2738} (\bibinfo{year}{2013}).

\bibitem[{\citenamefont{Wineland et~al.}(1992)\citenamefont{Wineland,
  Bollinger, Itano, Moore, and Heinzen}}]{wineland1992dj}
\bibinfo{author}{\bibfnamefont{D.~J.} \bibnamefont{Wineland}},
  \bibinfo{author}{\bibfnamefont{J.~J.} \bibnamefont{Bollinger}},
  \bibinfo{author}{\bibfnamefont{W.~M.} \bibnamefont{Itano}},
  \bibinfo{author}{\bibfnamefont{F.~L.} \bibnamefont{Moore}}, \bibnamefont{and}
  \bibinfo{author}{\bibfnamefont{D.~J.} \bibnamefont{Heinzen}},
  \bibinfo{journal}{Phys. Rev. A} \textbf{\bibinfo{volume}{46}},
  \bibinfo{pages}{R6797(R)} (\bibinfo{year}{1992}).

\bibitem[{\citenamefont{Huelga et~al.}(1997)\citenamefont{Huelga, Macchiavello,
  Pellizzari, Ekert, Plenio, and Cirac}}]{huelga1997improvement}
\bibinfo{author}{\bibfnamefont{S.~F.} \bibnamefont{Huelga}},
  \bibinfo{author}{\bibfnamefont{C.}~\bibnamefont{Macchiavello}},
  \bibinfo{author}{\bibfnamefont{T.}~\bibnamefont{Pellizzari}},
  \bibinfo{author}{\bibfnamefont{A.~K.} \bibnamefont{Ekert}},
  \bibinfo{author}{\bibfnamefont{M.~B.} \bibnamefont{Plenio}},
  \bibnamefont{and} \bibinfo{author}{\bibfnamefont{J.~I.} \bibnamefont{Cirac}},
  \bibinfo{journal}{Phys. Rev. Lett.} \textbf{\bibinfo{volume}{79}},
  \bibinfo{pages}{3865} (\bibinfo{year}{1997}).

\bibitem[{\citenamefont{Matsuzaki et~al.}(2011)\citenamefont{Matsuzaki,
  Benjamin, and Fitzsimons}}]{matsuzaki2011magnetic}
\bibinfo{author}{\bibfnamefont{Y.}~\bibnamefont{Matsuzaki}},
  \bibinfo{author}{\bibfnamefont{S.~C.} \bibnamefont{Benjamin}},
  \bibnamefont{and}
  \bibinfo{author}{\bibfnamefont{J.}~\bibnamefont{Fitzsimons}},
  \bibinfo{journal}{Phys. Rev. A} \textbf{\bibinfo{volume}{84}},
  \bibinfo{pages}{012103} (\bibinfo{year}{2011}).

\bibitem[{\citenamefont{Chin et~al.}(2012)\citenamefont{Chin, Huelga, and
  Plenio}}]{chin2012quantum}
\bibinfo{author}{\bibfnamefont{A.~W.} \bibnamefont{Chin}},
  \bibinfo{author}{\bibfnamefont{S.~F.} \bibnamefont{Huelga}},
  \bibnamefont{and} \bibinfo{author}{\bibfnamefont{M.~B.}
  \bibnamefont{Plenio}}, \bibinfo{journal}{Phys. Rev. Lett.}
  \textbf{\bibinfo{volume}{109}}, \bibinfo{pages}{233601}
  (\bibinfo{year}{2012}).

\bibitem[{\citenamefont{Kessler et~al.}(2014)\citenamefont{Kessler, Lovchinsky,
  Sushkov, and Lukin}}]{kessler2014quantum}
\bibinfo{author}{\bibfnamefont{E.~M.} \bibnamefont{Kessler}},
  \bibinfo{author}{\bibfnamefont{I.}~\bibnamefont{Lovchinsky}},
  \bibinfo{author}{\bibfnamefont{A.~O.} \bibnamefont{Sushkov}},
  \bibnamefont{and} \bibinfo{author}{\bibfnamefont{M.~D.} \bibnamefont{Lukin}},
  \bibinfo{journal}{Phys. Rev. Lett.} \textbf{\bibinfo{volume}{112}},
  \bibinfo{pages}{150802} (\bibinfo{year}{2014}).

\bibitem[{\citenamefont{D{\"u}r et~al.}(2014)\citenamefont{D{\"u}r,
  Skotiniotis, Fr\"{o}wis, and Kraus}}]{dur2014improved}
\bibinfo{author}{\bibfnamefont{W.}~\bibnamefont{D{\"u}r}},
  \bibinfo{author}{\bibfnamefont{M.}~\bibnamefont{Skotiniotis}},
  \bibinfo{author}{\bibfnamefont{F.}~\bibnamefont{Fr\"{o}wis}},
  \bibnamefont{and} \bibinfo{author}{\bibfnamefont{B.}~\bibnamefont{Kraus}},
  \bibinfo{journal}{Phys. Rev. Lett.} \textbf{\bibinfo{volume}{112}},
  \bibinfo{pages}{080801} (\bibinfo{year}{2014}).

\bibitem[{\citenamefont{Arrad et~al.}(2014)\citenamefont{Arrad, Vinkler,
  Aharonov, and Retzker}}]{arrad2014increasing}
\bibinfo{author}{\bibfnamefont{G.}~\bibnamefont{Arrad}},
  \bibinfo{author}{\bibfnamefont{Y.}~\bibnamefont{Vinkler}},
  \bibinfo{author}{\bibfnamefont{D.}~\bibnamefont{Aharonov}}, \bibnamefont{and}
  \bibinfo{author}{\bibfnamefont{A.}~\bibnamefont{Retzker}},
  \bibinfo{journal}{Phys. Rev. Lett.} \textbf{\bibinfo{volume}{112}},
  \bibinfo{pages}{150801} (\bibinfo{year}{2014}).

\bibitem[{\citenamefont{Herrera-Mart{\'\i}
  et~al.}(2015)\citenamefont{Herrera-Mart{\'\i}, Gefen, Aharonov, Katz, and
  Retzker}}]{herrera2015quantum}
\bibinfo{author}{\bibfnamefont{D.~A.} \bibnamefont{Herrera-Mart{\'\i}}},
  \bibinfo{author}{\bibfnamefont{T.}~\bibnamefont{Gefen}},
  \bibinfo{author}{\bibfnamefont{D.}~\bibnamefont{Aharonov}},
  \bibinfo{author}{\bibfnamefont{N.}~\bibnamefont{Katz}}, \bibnamefont{and}
  \bibinfo{author}{\bibfnamefont{A.}~\bibnamefont{Retzker}},
  \bibinfo{journal}{Phys. Rev. Lett.} \textbf{\bibinfo{volume}{115}},
  \bibinfo{pages}{200501} (\bibinfo{year}{2015}).

\bibitem[{\citenamefont{Unden~{\it et al.}}(2016)}]{unden2016quantum}
\bibinfo{author}{\bibfnamefont{T.}~\bibnamefont{Unden~{\it et al.}}},
  \bibinfo{journal}{Phys. Rev. Lett.} \textbf{\bibinfo{volume}{116}},
  \bibinfo{pages}{230502} (\bibinfo{year}{2016}).

\bibitem[{\citenamefont{Matsuzaki and Benjamin}(2017)}]{matsuzaki2017magnetic}
\bibinfo{author}{\bibfnamefont{Y.}~\bibnamefont{Matsuzaki}} \bibnamefont{and}
  \bibinfo{author}{\bibfnamefont{S.}~\bibnamefont{Benjamin}},
  \bibinfo{journal}{Phys. Rev. A} \textbf{\bibinfo{volume}{95}},
  \bibinfo{pages}{032303} (\bibinfo{year}{2017}).

\bibitem[{\citenamefont{Higgins et~al.}(2007)\citenamefont{Higgins, Berry,
  Bartlett, Wiseman, and Pryde}}]{higgins2007entanglement}
\bibinfo{author}{\bibfnamefont{B.~L.} \bibnamefont{Higgins}},
  \bibinfo{author}{\bibfnamefont{D.~W.} \bibnamefont{Berry}},
  \bibinfo{author}{\bibfnamefont{S.~D.} \bibnamefont{Bartlett}},
  \bibinfo{author}{\bibfnamefont{H.~M.} \bibnamefont{Wiseman}},
  \bibnamefont{and} \bibinfo{author}{\bibfnamefont{G.~J.} \bibnamefont{Pryde}},
  \bibinfo{journal}{Nature (London)} \textbf{\bibinfo{volume}{450}},
  \bibinfo{pages}{393} (\bibinfo{year}{2007}).

\bibitem[{\citenamefont{Waldherr et~al.}(2012)\citenamefont{Waldherr, Beck,
  Neumann, Said, Nitsche, Markham, Twitchen, Twamley, Jelezko, and
  Wrachtrup}}]{waldherr2012high}
\bibinfo{author}{\bibfnamefont{G.}~\bibnamefont{Waldherr}},
  \bibinfo{author}{\bibfnamefont{J.}~\bibnamefont{Beck}},
  \bibinfo{author}{\bibfnamefont{P.}~\bibnamefont{Neumann}},
  \bibinfo{author}{\bibfnamefont{R.}~\bibnamefont{Said}},
  \bibinfo{author}{\bibfnamefont{M.}~\bibnamefont{Nitsche}},
  \bibinfo{author}{\bibfnamefont{M.}~\bibnamefont{Markham}},
  \bibinfo{author}{\bibfnamefont{D.}~\bibnamefont{Twitchen}},
  \bibinfo{author}{\bibfnamefont{J.}~\bibnamefont{Twamley}},
  \bibinfo{author}{\bibfnamefont{F.}~\bibnamefont{Jelezko}}, \bibnamefont{and}
  \bibinfo{author}{\bibfnamefont{J.}~\bibnamefont{Wrachtrup}},
  \bibinfo{journal}{Nat. Nanotechnol.} \textbf{\bibinfo{volume}{7}},
  \bibinfo{pages}{105} (\bibinfo{year}{2012}).

\bibitem[{\citenamefont{Nakayama et~al.}(2015)\citenamefont{Nakayama, Soeda,
  and Murao}}]{nakayama2015quantum}
\bibinfo{author}{\bibfnamefont{S.}~\bibnamefont{Nakayama}},
  \bibinfo{author}{\bibfnamefont{A.}~\bibnamefont{Soeda}}, \bibnamefont{and}
  \bibinfo{author}{\bibfnamefont{M.}~\bibnamefont{Murao}},
  \bibinfo{journal}{Phys. Rev. Lett.} \textbf{\bibinfo{volume}{114}},
  \bibinfo{pages}{190501} (\bibinfo{year}{2015}).

\bibitem[{\citenamefont{Matsuzaki et~al.}(2017)\citenamefont{Matsuzaki,
  Nakayama, Soeda, Saito, and Murao}}]{matsuzaki2017projective}
\bibinfo{author}{\bibfnamefont{Y.}~\bibnamefont{Matsuzaki}},
  \bibinfo{author}{\bibfnamefont{S.}~\bibnamefont{Nakayama}},
  \bibinfo{author}{\bibfnamefont{A.}~\bibnamefont{Soeda}},
  \bibinfo{author}{\bibfnamefont{S.}~\bibnamefont{Saito}}, \bibnamefont{and}
  \bibinfo{author}{\bibfnamefont{M.}~\bibnamefont{Murao}},
  \bibinfo{journal}{Phys. Rev. A} \textbf{\bibinfo{volume}{95}},
  \bibinfo{pages}{062106} (\bibinfo{year}{2017}).

\bibitem[{\citenamefont{Komar et~al.}(2014)\citenamefont{Komar, Kessler,
  Bishof, Jiang, S{\o}rensen, Ye, and Lukin}}]{komar2014quantum}
\bibinfo{author}{\bibfnamefont{P.}~\bibnamefont{Komar}},
  \bibinfo{author}{\bibfnamefont{E.~M.} \bibnamefont{Kessler}},
  \bibinfo{author}{\bibfnamefont{M.}~\bibnamefont{Bishof}},
  \bibinfo{author}{\bibfnamefont{L.}~\bibnamefont{Jiang}},
  \bibinfo{author}{\bibfnamefont{A.~S.} \bibnamefont{S{\o}rensen}},
  \bibinfo{author}{\bibfnamefont{J.}~\bibnamefont{Ye}}, \bibnamefont{and}
  \bibinfo{author}{\bibfnamefont{M.~D.} \bibnamefont{Lukin}},
  \bibinfo{journal}{Nat. Phys.} \textbf{\bibinfo{volume}{10}},
  \bibinfo{pages}{582} (\bibinfo{year}{2014}).

\bibitem[{\citenamefont{Eldredge et~al.}(2018)\citenamefont{Eldredge,
  Foss-Feig, Gross, Rolston, and Gorshkov}}]{eldredge2018optimal}
\bibinfo{author}{\bibfnamefont{Z.}~\bibnamefont{Eldredge}},
  \bibinfo{author}{\bibfnamefont{M.}~\bibnamefont{Foss-Feig}},
  \bibinfo{author}{\bibfnamefont{J.~A.} \bibnamefont{Gross}},
  \bibinfo{author}{\bibfnamefont{S.~L.} \bibnamefont{Rolston}},
  \bibnamefont{and} \bibinfo{author}{\bibfnamefont{A.~V.}
  \bibnamefont{Gorshkov}}, \bibinfo{journal}{Phys. Rev. A}
  \textbf{\bibinfo{volume}{97}}, \bibinfo{pages}{042337}
  (\bibinfo{year}{2018}).

\bibitem[{\citenamefont{Proctor et~al.}(2018)\citenamefont{Proctor, Knott, and
  Dunningham}}]{proctor2018multiparameter}
\bibinfo{author}{\bibfnamefont{T.~J.} \bibnamefont{Proctor}},
  \bibinfo{author}{\bibfnamefont{P.~A.} \bibnamefont{Knott}}, \bibnamefont{and}
  \bibinfo{author}{\bibfnamefont{J.~A.} \bibnamefont{Dunningham}},
  \bibinfo{journal}{Phys. Rev. Lett.} \textbf{\bibinfo{volume}{120}},
  \bibinfo{pages}{080501} (\bibinfo{year}{2018}).

\bibitem[{\citenamefont{Lidar and Brun}(2013)}]{lidar2013quantum}
\bibinfo{author}{\bibfnamefont{D.~A.} \bibnamefont{Lidar}} \bibnamefont{and}
  \bibinfo{author}{\bibfnamefont{T.~A.} \bibnamefont{Brun}},
  \emph{\bibinfo{title}{Quantum error correction}}
  (\bibinfo{publisher}{Cambridge University Press, Cambridge},
  \bibinfo{year}{2013}).

\bibitem[{\citenamefont{Kitaev}(1996)}]{kitaev1997}
\bibinfo{author}{\bibfnamefont{A.~Y.} \bibnamefont{Kitaev}},
  \bibinfo{journal}{Electronic Colloquium on Computational Complexity}
  \textbf{\bibinfo{volume}{3}}, \bibinfo{pages}{3} (\bibinfo{year}{1996}).

\bibitem[{\citenamefont{Sasaki~{\it et al.}}(2011)}]{sasaki2011field}
\bibinfo{author}{\bibfnamefont{M.}~\bibnamefont{Sasaki~{\it et al.}}},
  \bibinfo{journal}{Optics express} \textbf{\bibinfo{volume}{19}},
  \bibinfo{pages}{10387} (\bibinfo{year}{2011}).

\bibitem[{\citenamefont{Wang~{\it et al.}}(2013)}]{wang2013direct}
\bibinfo{author}{\bibfnamefont{J.-Y.} \bibnamefont{Wang~{\it et al.}}},
  \bibinfo{journal}{Nat. Photon.} \textbf{\bibinfo{volume}{7}},
  \bibinfo{pages}{387} (\bibinfo{year}{2013}).

\bibitem[{\citenamefont{Giovannetti et~al.}(2001)\citenamefont{Giovannetti,
  Lloyd, and Maccone}}]{giovannetti2001quantum}
\bibinfo{author}{\bibfnamefont{V.}~\bibnamefont{Giovannetti}},
  \bibinfo{author}{\bibfnamefont{S.}~\bibnamefont{Lloyd}}, \bibnamefont{and}
  \bibinfo{author}{\bibfnamefont{L.}~\bibnamefont{Maccone}},
  \bibinfo{journal}{Nature (London)} \textbf{\bibinfo{volume}{412}},
  \bibinfo{pages}{417} (\bibinfo{year}{2001}).

\bibitem[{\citenamefont{Giovannetti
  et~al.}(2002{\natexlab{a}})\citenamefont{Giovannetti, Lloyd, and
  Maccone}}]{giovannetti2002quantum}
\bibinfo{author}{\bibfnamefont{V.}~\bibnamefont{Giovannetti}},
  \bibinfo{author}{\bibfnamefont{S.}~\bibnamefont{Lloyd}}, \bibnamefont{and}
  \bibinfo{author}{\bibfnamefont{L.}~\bibnamefont{Maccone}},
  \bibinfo{journal}{J. Opt. B: Quantum Semiclassical Opt.}
  \textbf{\bibinfo{volume}{4}}, \bibinfo{pages}{S413}
  (\bibinfo{year}{2002}{\natexlab{a}}).

\bibitem[{\citenamefont{Giovannetti
  et~al.}(2002{\natexlab{b}})\citenamefont{Giovannetti, Lloyd, and
  Maccone}}]{giovannetti2002positioning}
\bibinfo{author}{\bibfnamefont{V.}~\bibnamefont{Giovannetti}},
  \bibinfo{author}{\bibfnamefont{S.}~\bibnamefont{Lloyd}}, \bibnamefont{and}
  \bibinfo{author}{\bibfnamefont{L.}~\bibnamefont{Maccone}},
  \bibinfo{journal}{Phys. Rev. A} \textbf{\bibinfo{volume}{65}},
  \bibinfo{pages}{022309} (\bibinfo{year}{2002}{\natexlab{b}}).

\bibitem[{\citenamefont{Chiribella et~al.}(2005)\citenamefont{Chiribella,
  D'Ariano, and Sacchi}}]{chiribella2005optimal}
\bibinfo{author}{\bibfnamefont{G.}~\bibnamefont{Chiribella}},
  \bibinfo{author}{\bibfnamefont{G.~M.} \bibnamefont{D'Ariano}},
  \bibnamefont{and} \bibinfo{author}{\bibfnamefont{M.~F.}
  \bibnamefont{Sacchi}}, \bibinfo{journal}{Phys. Rev. A}
  \textbf{\bibinfo{volume}{72}}, \bibinfo{pages}{042338}
  (\bibinfo{year}{2005}).

\bibitem[{\citenamefont{Chiribella et~al.}(2007)\citenamefont{Chiribella,
  Maccone, and Perinotti}}]{chiribella2007secret}
\bibinfo{author}{\bibfnamefont{G.}~\bibnamefont{Chiribella}},
  \bibinfo{author}{\bibfnamefont{L.}~\bibnamefont{Maccone}}, \bibnamefont{and}
  \bibinfo{author}{\bibfnamefont{P.}~\bibnamefont{Perinotti}},
  \bibinfo{journal}{Phys. Rev. Lett.} \textbf{\bibinfo{volume}{98}},
  \bibinfo{pages}{120501} (\bibinfo{year}{2007}).

\bibitem[{\citenamefont{Huang et~al.}(2017)\citenamefont{Huang, Macchiavello,
  and Maccone}}]{huang2017cryptographic}
\bibinfo{author}{\bibfnamefont{Z.}~\bibnamefont{Huang}},
  \bibinfo{author}{\bibfnamefont{C.}~\bibnamefont{Macchiavello}},
  \bibnamefont{and} \bibinfo{author}{\bibfnamefont{L.}~\bibnamefont{Maccone}},
  \bibinfo{journal}{arXiv:1706.03894}  (\bibinfo{year}{2017}).

\bibitem[{\citenamefont{Xie et~al.}(2018)\citenamefont{Xie, Xu, Chen, and
  Wang}}]{xie2018high}
\bibinfo{author}{\bibfnamefont{D.}~\bibnamefont{Xie}},
  \bibinfo{author}{\bibfnamefont{C.}~\bibnamefont{Xu}},
  \bibinfo{author}{\bibfnamefont{J.}~\bibnamefont{Chen}}, \bibnamefont{and}
  \bibinfo{author}{\bibfnamefont{A.~M.} \bibnamefont{Wang}},
  \bibinfo{journal}{Quant. Info. Proc.} \textbf{\bibinfo{volume}{17}},
  \bibinfo{pages}{116} (\bibinfo{year}{2018}).

\bibitem[{\citenamefont{Bennett et~al.}(1993)\citenamefont{Bennett, Brassard,
  Cr{\'e}peau, Jozsa, Peres, and Wootters}}]{bennett1993teleporting}
\bibinfo{author}{\bibfnamefont{C.~H.} \bibnamefont{Bennett}},
  \bibinfo{author}{\bibfnamefont{G.}~\bibnamefont{Brassard}},
  \bibinfo{author}{\bibfnamefont{C.}~\bibnamefont{Cr{\'e}peau}},
  \bibinfo{author}{\bibfnamefont{R.}~\bibnamefont{Jozsa}},
  \bibinfo{author}{\bibfnamefont{A.}~\bibnamefont{Peres}}, \bibnamefont{and}
  \bibinfo{author}{\bibfnamefont{W.~K.} \bibnamefont{Wootters}},
  \bibinfo{journal}{Phys. Rev. Lett.} \textbf{\bibinfo{volume}{70}},
  \bibinfo{pages}{1895} (\bibinfo{year}{1993}).

\bibitem[{\citenamefont{Bouwmeester et~al.}(1997)\citenamefont{Bouwmeester,
  Pan, Mattle, Eibl, Weinfurter, and Zeilinger}}]{bouwmeester1997experimental}
\bibinfo{author}{\bibfnamefont{D.}~\bibnamefont{Bouwmeester}},
  \bibinfo{author}{\bibfnamefont{J.-W.} \bibnamefont{Pan}},
  \bibinfo{author}{\bibfnamefont{K.}~\bibnamefont{Mattle}},
  \bibinfo{author}{\bibfnamefont{M.}~\bibnamefont{Eibl}},
  \bibinfo{author}{\bibfnamefont{H.}~\bibnamefont{Weinfurter}},
  \bibnamefont{and}
  \bibinfo{author}{\bibfnamefont{A.}~\bibnamefont{Zeilinger}},
  \bibinfo{journal}{Nature (London)} \textbf{\bibinfo{volume}{390}},
  \bibinfo{pages}{575} (\bibinfo{year}{1997}).

\bibitem[{\citenamefont{Furusawa et~al.}(1998)\citenamefont{Furusawa,
  S{\o}rensen, Braunstein, Fuchs, Kimble, and
  Polzik}}]{furusawa1998unconditional}
\bibinfo{author}{\bibfnamefont{A.}~\bibnamefont{Furusawa}},
  \bibinfo{author}{\bibfnamefont{J.~L.} \bibnamefont{S{\o}rensen}},
  \bibinfo{author}{\bibfnamefont{S.~L.} \bibnamefont{Braunstein}},
  \bibinfo{author}{\bibfnamefont{C.~A.} \bibnamefont{Fuchs}},
  \bibinfo{author}{\bibfnamefont{H.~J.} \bibnamefont{Kimble}},
  \bibnamefont{and} \bibinfo{author}{\bibfnamefont{E.~S.}
  \bibnamefont{Polzik}}, \bibinfo{journal}{Science}
  \textbf{\bibinfo{volume}{282}}, \bibinfo{pages}{706} (\bibinfo{year}{1998}).

\bibitem[{\citenamefont{Julsgaard et~al.}(2004)\citenamefont{Julsgaard,
  Sherson, Cirac, Fiur{\'a}{\v{s}}ek, and Polzik}}]{julsgaard2004experimental}
\bibinfo{author}{\bibfnamefont{B.}~\bibnamefont{Julsgaard}},
  \bibinfo{author}{\bibfnamefont{J.}~\bibnamefont{Sherson}},
  \bibinfo{author}{\bibfnamefont{J.~I.} \bibnamefont{Cirac}},
  \bibinfo{author}{\bibfnamefont{J.}~\bibnamefont{Fiur{\'a}{\v{s}}ek}},
  \bibnamefont{and} \bibinfo{author}{\bibfnamefont{E.~S.}
  \bibnamefont{Polzik}}, \bibinfo{journal}{Nature (London)}
  \textbf{\bibinfo{volume}{432}}, \bibinfo{pages}{482} (\bibinfo{year}{2004}).

\bibitem[{\citenamefont{Hedges et~al.}(2010)\citenamefont{Hedges, Longdell, Li,
  and Sellars}}]{hedges2010efficient}
\bibinfo{author}{\bibfnamefont{M.~P.} \bibnamefont{Hedges}},
  \bibinfo{author}{\bibfnamefont{J.~J.} \bibnamefont{Longdell}},
  \bibinfo{author}{\bibfnamefont{Y.}~\bibnamefont{Li}}, \bibnamefont{and}
  \bibinfo{author}{\bibfnamefont{M.~J.} \bibnamefont{Sellars}},
  \bibinfo{journal}{Nature (London)} \textbf{\bibinfo{volume}{465}},
  \bibinfo{pages}{1052} (\bibinfo{year}{2010}).

\bibitem[{\citenamefont{Smithey et~al.}(1993)\citenamefont{Smithey, Beck,
  Raymer, and Faridani}}]{smithey1993measurement}
\bibinfo{author}{\bibfnamefont{D.~T.} \bibnamefont{Smithey}},
  \bibinfo{author}{\bibfnamefont{M.}~\bibnamefont{Beck}},
  \bibinfo{author}{\bibfnamefont{M.~G.} \bibnamefont{Raymer}},
  \bibnamefont{and} \bibinfo{author}{\bibfnamefont{A.}~\bibnamefont{Faridani}},
  \bibinfo{journal}{Phys. Rev. Lett.} \textbf{\bibinfo{volume}{70}},
  \bibinfo{pages}{1244} (\bibinfo{year}{1993}).

\bibitem[{\citenamefont{Hradil}(1997)}]{hradil1997quantum}
\bibinfo{author}{\bibfnamefont{Z.}~\bibnamefont{Hradil}},
  \bibinfo{journal}{Phys. Rev. A} \textbf{\bibinfo{volume}{55}},
  \bibinfo{pages}{R1561} (\bibinfo{year}{1997}).

\bibitem[{\citenamefont{Banaszek et~al.}(1999)\citenamefont{Banaszek, D'Ariano,
  Paris, and Sacchi}}]{banaszek1999maximum}
\bibinfo{author}{\bibfnamefont{K.}~\bibnamefont{Banaszek}},
  \bibinfo{author}{\bibfnamefont{G.~M.} \bibnamefont{D'Ariano}},
  \bibinfo{author}{\bibfnamefont{M.~G.~A.} \bibnamefont{Paris}},
  \bibnamefont{and} \bibinfo{author}{\bibfnamefont{M.~F.}
  \bibnamefont{Sacchi}}, \bibinfo{journal}{Phys. Rev. A}
  \textbf{\bibinfo{volume}{61}}, \bibinfo{pages}{010304}
  (\bibinfo{year}{1999}).

\bibitem[{\citenamefont{Poyatos et~al.}(1997)\citenamefont{Poyatos, Cirac, and
  Zoller}}]{poyatos1997complete}
\bibinfo{author}{\bibfnamefont{J.~F.} \bibnamefont{Poyatos}},
  \bibinfo{author}{\bibfnamefont{J.~I.} \bibnamefont{Cirac}}, \bibnamefont{and}
  \bibinfo{author}{\bibfnamefont{P.}~\bibnamefont{Zoller}},
  \bibinfo{journal}{Phys. Rev. Lett.} \textbf{\bibinfo{volume}{78}},
  \bibinfo{pages}{390} (\bibinfo{year}{1997}).

\bibitem[{\citenamefont{Chuang and Nielsen}(1997)}]{chuang1997prescription}
\bibinfo{author}{\bibfnamefont{I.~L.} \bibnamefont{Chuang}} \bibnamefont{and}
  \bibinfo{author}{\bibfnamefont{M.~A.} \bibnamefont{Nielsen}},
  \bibinfo{journal}{J. Mod. Opt.} \textbf{\bibinfo{volume}{44}},
  \bibinfo{pages}{2455} (\bibinfo{year}{1997}).

\bibitem[{\citenamefont{Nielsen and Chuang}(2000)}]{NC01b}
\bibinfo{author}{\bibfnamefont{M.~A.} \bibnamefont{Nielsen}} \bibnamefont{and}
  \bibinfo{author}{\bibfnamefont{I.~L.} \bibnamefont{Chuang}},
  \emph{\bibinfo{title}{Quantum Computation and Quantum Information}}
  (\bibinfo{publisher}{Cambridge University Press, Cambridge},
  \bibinfo{year}{2000}).

\bibitem[{\citenamefont{Takeuchi et~al.}(2018)\citenamefont{Takeuchi, Mantri,
  Morimae, Mizutani, and Fitzsimons}}]{takeuchi2018resource}
\bibinfo{author}{\bibfnamefont{Y.}~\bibnamefont{Takeuchi}},
  \bibinfo{author}{\bibfnamefont{A.}~\bibnamefont{Mantri}},
  \bibinfo{author}{\bibfnamefont{T.}~\bibnamefont{Morimae}},
  \bibinfo{author}{\bibfnamefont{A.}~\bibnamefont{Mizutani}}, \bibnamefont{and}
  \bibinfo{author}{\bibfnamefont{J.~F.} \bibnamefont{Fitzsimons}},
  \bibinfo{journal}{arXiv:1806.09138}  (\bibinfo{year}{2018}).

\bibitem[{\citenamefont{Hayashi and Morimae}(2015)}]{hayashi2015verifiable}
\bibinfo{author}{\bibfnamefont{M.}~\bibnamefont{Hayashi}} \bibnamefont{and}
  \bibinfo{author}{\bibfnamefont{T.}~\bibnamefont{Morimae}},
  \bibinfo{journal}{Phys. Rev. Lett.} \textbf{\bibinfo{volume}{115}},
  \bibinfo{pages}{220502} (\bibinfo{year}{2015}).

\bibitem[{\citenamefont{Markham and Krause}(2018)}]{markham2018simple}
\bibinfo{author}{\bibfnamefont{D.}~\bibnamefont{Markham}} \bibnamefont{and}
  \bibinfo{author}{\bibfnamefont{A.}~\bibnamefont{Krause}},
  \bibinfo{journal}{arXiv:1801.05057}  (\bibinfo{year}{2018}).

\bibitem[{\citenamefont{Taylor et~al.}(2008)\citenamefont{Taylor, Cappellaro,
  Childress, Jiang, Budker, Hemmer, Yacoby, Walsworth, and
  Lukin}}]{taylor2008high}
\bibinfo{author}{\bibfnamefont{J.}~\bibnamefont{Taylor}},
  \bibinfo{author}{\bibfnamefont{P.}~\bibnamefont{Cappellaro}},
  \bibinfo{author}{\bibfnamefont{L.}~\bibnamefont{Childress}},
  \bibinfo{author}{\bibfnamefont{L.}~\bibnamefont{Jiang}},
  \bibinfo{author}{\bibfnamefont{D.}~\bibnamefont{Budker}},
  \bibinfo{author}{\bibfnamefont{P.}~\bibnamefont{Hemmer}},
  \bibinfo{author}{\bibfnamefont{A.}~\bibnamefont{Yacoby}},
  \bibinfo{author}{\bibfnamefont{R.}~\bibnamefont{Walsworth}},
  \bibnamefont{and} \bibinfo{author}{\bibfnamefont{M.}~\bibnamefont{Lukin}},
  \bibinfo{journal}{Nat. Phys.} \textbf{\bibinfo{volume}{4}},
  \bibinfo{pages}{810} (\bibinfo{year}{2008}).

\bibitem[{\citenamefont{Collins}(1994)}]{Go01a}
\bibinfo{author}{\bibfnamefont{A.~T.} \bibnamefont{Collins}},
  \emph{\bibinfo{title}{Properties and Growth of Diamond, {\rm edited by G.
  Davies}}} (\bibinfo{publisher}{INSPEC, London}, \bibinfo{year}{1994}).

\bibitem[{\citenamefont{Gruber et~al.}(1997)\citenamefont{Gruber,
  Dr{\"a}benstedt, Tietz, Fleury, Wrachtrup, and
  Von~Borczyskowski}}]{gruber1997scanning}
\bibinfo{author}{\bibfnamefont{A.}~\bibnamefont{Gruber}},
  \bibinfo{author}{\bibfnamefont{A.}~\bibnamefont{Dr{\"a}benstedt}},
  \bibinfo{author}{\bibfnamefont{C.}~\bibnamefont{Tietz}},
  \bibinfo{author}{\bibfnamefont{L.}~\bibnamefont{Fleury}},
  \bibinfo{author}{\bibfnamefont{J.}~\bibnamefont{Wrachtrup}},
  \bibnamefont{and}
  \bibinfo{author}{\bibfnamefont{C.}~\bibnamefont{Von~Borczyskowski}},
  \bibinfo{journal}{Science} \textbf{\bibinfo{volume}{276}},
  \bibinfo{pages}{2012} (\bibinfo{year}{1997}).

\bibitem[{\citenamefont{Jelezko et~al.}(2002)\citenamefont{Jelezko, Popa,
  Gruber, Tietz, Wrachtrup, Nizovtsev, and Kilin}}]{jelezko2002single}
\bibinfo{author}{\bibfnamefont{F.}~\bibnamefont{Jelezko}},
  \bibinfo{author}{\bibfnamefont{I.}~\bibnamefont{Popa}},
  \bibinfo{author}{\bibfnamefont{A.}~\bibnamefont{Gruber}},
  \bibinfo{author}{\bibfnamefont{C.}~\bibnamefont{Tietz}},
  \bibinfo{author}{\bibfnamefont{J.}~\bibnamefont{Wrachtrup}},
  \bibinfo{author}{\bibfnamefont{A.}~\bibnamefont{Nizovtsev}},
  \bibnamefont{and} \bibinfo{author}{\bibfnamefont{S.}~\bibnamefont{Kilin}},
  \bibinfo{journal}{Appl. Phys. Lett.} \textbf{\bibinfo{volume}{81}},
  \bibinfo{pages}{2160} (\bibinfo{year}{2002}).

\bibitem[{\citenamefont{Jelezko et~al.}(2004)\citenamefont{Jelezko, Gaebel,
  Popa, Gruber, and Wrachtrup}}]{JGPGW01a}
\bibinfo{author}{\bibfnamefont{F.}~\bibnamefont{Jelezko}},
  \bibinfo{author}{\bibfnamefont{T.}~\bibnamefont{Gaebel}},
  \bibinfo{author}{\bibfnamefont{I.}~\bibnamefont{Popa}},
  \bibinfo{author}{\bibfnamefont{A.}~\bibnamefont{Gruber}}, \bibnamefont{and}
  \bibinfo{author}{\bibfnamefont{J.}~\bibnamefont{Wrachtrup}},
  \bibinfo{journal}{Phys. Rev. Lett.} \textbf{\bibinfo{volume}{92}},
  \bibinfo{pages}{076401} (\bibinfo{year}{2004}).

\bibitem[{\citenamefont{Jiang~{\it et al.}}(2009)}]{jiang2009repetitive}
\bibinfo{author}{\bibfnamefont{L.}~\bibnamefont{Jiang~{\it et al.}}},
  \bibinfo{journal}{Science} \textbf{\bibinfo{volume}{326}},
  \bibinfo{pages}{267} (\bibinfo{year}{2009}).

\bibitem[{\citenamefont{Barrett and Kok}(2005)}]{Barrett:2005p363}
\bibinfo{author}{\bibfnamefont{S.~D.} \bibnamefont{Barrett}} \bibnamefont{and}
  \bibinfo{author}{\bibfnamefont{P.}~\bibnamefont{Kok}},
  \bibinfo{journal}{Phys. Rev. A} \textbf{\bibinfo{volume}{71}},
  \bibinfo{pages}{060310(R)} (\bibinfo{year}{2005}).

\bibitem[{\citenamefont{Childress et~al.}(2006)\citenamefont{Childress, Taylor,
  S\o{}rensen, and Lukin}}]{childress2006fault}
\bibinfo{author}{\bibfnamefont{L.}~\bibnamefont{Childress}},
  \bibinfo{author}{\bibfnamefont{J.~M.} \bibnamefont{Taylor}},
  \bibinfo{author}{\bibfnamefont{A.~S.} \bibnamefont{S\o{}rensen}},
  \bibnamefont{and} \bibinfo{author}{\bibfnamefont{M.~D.} \bibnamefont{Lukin}},
  \bibinfo{journal}{Phys. Rev. Lett.} \textbf{\bibinfo{volume}{96}},
  \bibinfo{pages}{070504} (\bibinfo{year}{2006}).

\bibitem[{\citenamefont{Nemoto et~al.}(2014)\citenamefont{Nemoto, Trupke,
  Devitt, Stephens, Scharfenberger, Buczak, N{\"o}bauer, Everitt, Schmiedmayer,
  and Munro}}]{nemoto2014photonic}
\bibinfo{author}{\bibfnamefont{K.}~\bibnamefont{Nemoto}},
  \bibinfo{author}{\bibfnamefont{M.}~\bibnamefont{Trupke}},
  \bibinfo{author}{\bibfnamefont{S.~J.} \bibnamefont{Devitt}},
  \bibinfo{author}{\bibfnamefont{A.~M.} \bibnamefont{Stephens}},
  \bibinfo{author}{\bibfnamefont{B.}~\bibnamefont{Scharfenberger}},
  \bibinfo{author}{\bibfnamefont{K.}~\bibnamefont{Buczak}},
  \bibinfo{author}{\bibfnamefont{T.}~\bibnamefont{N{\"o}bauer}},
  \bibinfo{author}{\bibfnamefont{M.~S.} \bibnamefont{Everitt}},
  \bibinfo{author}{\bibfnamefont{J.}~\bibnamefont{Schmiedmayer}},
  \bibnamefont{and} \bibinfo{author}{\bibfnamefont{W.~J.} \bibnamefont{Munro}},
  \bibinfo{journal}{Phys. Rev. X} \textbf{\bibinfo{volume}{4}},
  \bibinfo{pages}{031022} (\bibinfo{year}{2014}).

\bibitem[{\citenamefont{Nemoto et~al.}(2016)\citenamefont{Nemoto, Trupke,
  Devitt, Scharfenberger, Buczak, Schmiedmayer, and
  Munro}}]{nemoto2016photonic}
\bibinfo{author}{\bibfnamefont{K.}~\bibnamefont{Nemoto}},
  \bibinfo{author}{\bibfnamefont{M.}~\bibnamefont{Trupke}},
  \bibinfo{author}{\bibfnamefont{S.~J.} \bibnamefont{Devitt}},
  \bibinfo{author}{\bibfnamefont{B.}~\bibnamefont{Scharfenberger}},
  \bibinfo{author}{\bibfnamefont{K.}~\bibnamefont{Buczak}},
  \bibinfo{author}{\bibfnamefont{J.}~\bibnamefont{Schmiedmayer}},
  \bibnamefont{and} \bibinfo{author}{\bibfnamefont{W.~J.} \bibnamefont{Munro}},
  \bibinfo{journal}{Sci. Rep.} \textbf{\bibinfo{volume}{6}},
  \bibinfo{pages}{26284} (\bibinfo{year}{2016}).

\bibitem[{\citenamefont{Togan~{\it et al.}}(2010)}]{togan2010quantum}
\bibinfo{author}{\bibfnamefont{E.}~\bibnamefont{Togan~{\it et al.}}},
  \bibinfo{journal}{Nature (London)} \textbf{\bibinfo{volume}{466}},
  \bibinfo{pages}{730} (\bibinfo{year}{2010}).

\bibitem[{\citenamefont{Rondin et~al.}(2012)\citenamefont{Rondin, Tetienne,
  Spinicelli, Dal~Savio, Karrai, Dantelle, Thiaville, Rohart, Roch, and
  Jacques}}]{rondin2012nanoscale}
\bibinfo{author}{\bibfnamefont{L.}~\bibnamefont{Rondin}},
  \bibinfo{author}{\bibfnamefont{J.~P.} \bibnamefont{Tetienne}},
  \bibinfo{author}{\bibfnamefont{P.}~\bibnamefont{Spinicelli}},
  \bibinfo{author}{\bibfnamefont{C.}~\bibnamefont{Dal~Savio}},
  \bibinfo{author}{\bibfnamefont{K.}~\bibnamefont{Karrai}},
  \bibinfo{author}{\bibfnamefont{G.}~\bibnamefont{Dantelle}},
  \bibinfo{author}{\bibfnamefont{A.}~\bibnamefont{Thiaville}},
  \bibinfo{author}{\bibfnamefont{S.}~\bibnamefont{Rohart}},
  \bibinfo{author}{\bibfnamefont{J.~F.} \bibnamefont{Roch}}, \bibnamefont{and}
  \bibinfo{author}{\bibfnamefont{V.}~\bibnamefont{Jacques}},
  \bibinfo{journal}{Appl. Phys. Lett.} \textbf{\bibinfo{volume}{100}},
  \bibinfo{pages}{153118} (\bibinfo{year}{2012}).

\bibitem[{\citenamefont{Cuche et~al.}(2009)\citenamefont{Cuche, Drezet,
  Sonnefraud, Faklaris, Treussart, Roch, and Huant}}]{cuche2009near}
\bibinfo{author}{\bibfnamefont{A.}~\bibnamefont{Cuche}},
  \bibinfo{author}{\bibfnamefont{A.}~\bibnamefont{Drezet}},
  \bibinfo{author}{\bibfnamefont{Y.}~\bibnamefont{Sonnefraud}},
  \bibinfo{author}{\bibfnamefont{O.}~\bibnamefont{Faklaris}},
  \bibinfo{author}{\bibfnamefont{F.}~\bibnamefont{Treussart}},
  \bibinfo{author}{\bibfnamefont{J.-F.} \bibnamefont{Roch}}, \bibnamefont{and}
  \bibinfo{author}{\bibfnamefont{S.}~\bibnamefont{Huant}},
  \bibinfo{journal}{Opt. Express} \textbf{\bibinfo{volume}{17}},
  \bibinfo{pages}{19969} (\bibinfo{year}{2009}).

\bibitem[{\citenamefont{Maletinsky et~al.}(2012)\citenamefont{Maletinsky, Hong,
  Grinolds, Hausmann, Lukin, Walsworth, Loncar, and
  Yacoby}}]{maletinsky2012robust}
\bibinfo{author}{\bibfnamefont{P.}~\bibnamefont{Maletinsky}},
  \bibinfo{author}{\bibfnamefont{S.}~\bibnamefont{Hong}},
  \bibinfo{author}{\bibfnamefont{M.~S.} \bibnamefont{Grinolds}},
  \bibinfo{author}{\bibfnamefont{B.}~\bibnamefont{Hausmann}},
  \bibinfo{author}{\bibfnamefont{M.~D.} \bibnamefont{Lukin}},
  \bibinfo{author}{\bibfnamefont{R.~L.} \bibnamefont{Walsworth}},
  \bibinfo{author}{\bibfnamefont{M.}~\bibnamefont{Loncar}}, \bibnamefont{and}
  \bibinfo{author}{\bibfnamefont{A.}~\bibnamefont{Yacoby}},
  \bibinfo{journal}{Nat. Nanotechnol.} \textbf{\bibinfo{volume}{7}},
  \bibinfo{pages}{320} (\bibinfo{year}{2012}).

\bibitem[{\citenamefont{Twamley and
  Barrett}(2010)}]{twamley2010superconducting}
\bibinfo{author}{\bibfnamefont{J.}~\bibnamefont{Twamley}} \bibnamefont{and}
  \bibinfo{author}{\bibfnamefont{S.~D.} \bibnamefont{Barrett}},
  \bibinfo{journal}{Phys. Rev. B} \textbf{\bibinfo{volume}{81}},
  \bibinfo{pages}{241202} (\bibinfo{year}{2010}).

\bibitem[{\citenamefont{Marcos et~al.}(2010)\citenamefont{Marcos, Wubs, Taylor,
  Aguado, Lukin, and S{\o}rensen}}]{marcos2010coupling}
\bibinfo{author}{\bibfnamefont{D.}~\bibnamefont{Marcos}},
  \bibinfo{author}{\bibfnamefont{M.}~\bibnamefont{Wubs}},
  \bibinfo{author}{\bibfnamefont{J.~M.} \bibnamefont{Taylor}},
  \bibinfo{author}{\bibfnamefont{R.}~\bibnamefont{Aguado}},
  \bibinfo{author}{\bibfnamefont{M.~D.} \bibnamefont{Lukin}}, \bibnamefont{and}
  \bibinfo{author}{\bibfnamefont{A.~S.} \bibnamefont{S{\o}rensen}},
  \bibinfo{journal}{Phys. Rev. Lett.} \textbf{\bibinfo{volume}{105}},
  \bibinfo{pages}{210501} (\bibinfo{year}{2010}).

\bibitem[{\citenamefont{Zhu~{\it et al.}}(2011)}]{zhu2011coherent}
\bibinfo{author}{\bibfnamefont{X.}~\bibnamefont{Zhu~{\it et al.}}},
  \bibinfo{journal}{Nature (London)} \textbf{\bibinfo{volume}{478}},
  \bibinfo{pages}{221} (\bibinfo{year}{2011}).

\bibitem[{\citenamefont{Zhu et~al.}(2014)\citenamefont{Zhu, Matsuzaki, Amsuss,
  Kakuyanagi, Shimo-Oka, Mizuochi, Nemoto, Munro, Semba, and
  Saito}}]{zhudark2014}
\bibinfo{author}{\bibfnamefont{X.}~\bibnamefont{Zhu}},
  \bibinfo{author}{\bibfnamefont{Y.}~\bibnamefont{Matsuzaki}},
  \bibinfo{author}{\bibfnamefont{R.}~\bibnamefont{Amsuss}},
  \bibinfo{author}{\bibfnamefont{K.}~\bibnamefont{Kakuyanagi}},
  \bibinfo{author}{\bibfnamefont{T.}~\bibnamefont{Shimo-Oka}},
  \bibinfo{author}{\bibfnamefont{N.}~\bibnamefont{Mizuochi}},
  \bibinfo{author}{\bibfnamefont{K.}~\bibnamefont{Nemoto}},
  \bibinfo{author}{\bibfnamefont{W.~J.} \bibnamefont{Munro}},
  \bibinfo{author}{\bibfnamefont{K.}~\bibnamefont{Semba}}, \bibnamefont{and}
  \bibinfo{author}{\bibfnamefont{S.}~\bibnamefont{Saito}},
  \bibinfo{journal}{Nat. Commun.} \textbf{\bibinfo{volume}{3424}},
  \bibinfo{pages}{4524} (\bibinfo{year}{2014}).

\bibitem[{\citenamefont{Matsuzaki~{\it et al.}}(2015)}]{matsuzaki2015improving}
\bibinfo{author}{\bibfnamefont{Y.}~\bibnamefont{Matsuzaki~{\it et al.}}},
  \bibinfo{journal}{Phys. Rev. Lett.} \textbf{\bibinfo{volume}{114}},
  \bibinfo{pages}{120501} (\bibinfo{year}{2015}).

\bibitem[{\citenamefont{Bylander et~al.}(2011)\citenamefont{Bylander,
  Gustavsson, Yan, Yoshihara, Harrabi, Fitch, Cory, Nakamura, Tsai, and
  Oliver}}]{bylander2011noise}
\bibinfo{author}{\bibfnamefont{J.}~\bibnamefont{Bylander}},
  \bibinfo{author}{\bibfnamefont{S.}~\bibnamefont{Gustavsson}},
  \bibinfo{author}{\bibfnamefont{F.}~\bibnamefont{Yan}},
  \bibinfo{author}{\bibfnamefont{F.}~\bibnamefont{Yoshihara}},
  \bibinfo{author}{\bibfnamefont{K.}~\bibnamefont{Harrabi}},
  \bibinfo{author}{\bibfnamefont{G.}~\bibnamefont{Fitch}},
  \bibinfo{author}{\bibfnamefont{D.~G.} \bibnamefont{Cory}},
  \bibinfo{author}{\bibfnamefont{Y.}~\bibnamefont{Nakamura}},
  \bibinfo{author}{\bibfnamefont{J.~S.} \bibnamefont{Tsai}}, \bibnamefont{and}
  \bibinfo{author}{\bibfnamefont{W.~D.} \bibnamefont{Oliver}},
  \bibinfo{journal}{Nat. Phys.} \textbf{\bibinfo{volume}{7}},
  \bibinfo{pages}{565} (\bibinfo{year}{2011}).

\bibitem[{\citenamefont{Clarke and Wilhelm}(2007)}]{ClarkeWilhelm01a}
\bibinfo{author}{\bibfnamefont{J.}~\bibnamefont{Clarke}} \bibnamefont{and}
  \bibinfo{author}{\bibfnamefont{F.~K.} \bibnamefont{Wilhelm}},
  \bibinfo{journal}{Nature (London)} \textbf{\bibinfo{volume}{453}},
  \bibinfo{pages}{1031} (\bibinfo{year}{2007}).

\bibitem[{\citenamefont{Bal et~al.}(2012)\citenamefont{Bal, Deng, Orgiazzi,
  Ong, and Lupascu}}]{bal2012ultrasensitive}
\bibinfo{author}{\bibfnamefont{M.}~\bibnamefont{Bal}},
  \bibinfo{author}{\bibfnamefont{C.}~\bibnamefont{Deng}},
  \bibinfo{author}{\bibfnamefont{J.-L.} \bibnamefont{Orgiazzi}},
  \bibinfo{author}{\bibfnamefont{F.}~\bibnamefont{Ong}}, \bibnamefont{and}
  \bibinfo{author}{\bibfnamefont{A.}~\bibnamefont{Lupascu}},
  \bibinfo{journal}{Nat. Commun.} \textbf{\bibinfo{volume}{3}},
  \bibinfo{pages}{1324} (\bibinfo{year}{2012}).

\bibitem[{\citenamefont{Saito et~al.}(2013)\citenamefont{Saito, Zhu,
  Ams{\"u}ss, Matsuzaki, Kakuyanagi, Shimo-Oka, Mizuochi, Nemoto, Munro, and
  Semba}}]{saito2013towards}
\bibinfo{author}{\bibfnamefont{S.}~\bibnamefont{Saito}},
  \bibinfo{author}{\bibfnamefont{X.}~\bibnamefont{Zhu}},
  \bibinfo{author}{\bibfnamefont{R.}~\bibnamefont{Ams{\"u}ss}},
  \bibinfo{author}{\bibfnamefont{Y.}~\bibnamefont{Matsuzaki}},
  \bibinfo{author}{\bibfnamefont{K.}~\bibnamefont{Kakuyanagi}},
  \bibinfo{author}{\bibfnamefont{T.}~\bibnamefont{Shimo-Oka}},
  \bibinfo{author}{\bibfnamefont{N.}~\bibnamefont{Mizuochi}},
  \bibinfo{author}{\bibfnamefont{K.}~\bibnamefont{Nemoto}},
  \bibinfo{author}{\bibfnamefont{W.~J.} \bibnamefont{Munro}}, \bibnamefont{and}
  \bibinfo{author}{\bibfnamefont{K.}~\bibnamefont{Semba}},
  \bibinfo{journal}{Phys. Rev. Lett.} \textbf{\bibinfo{volume}{111}},
  \bibinfo{pages}{107008} (\bibinfo{year}{2013}).

\bibitem[{\citenamefont{Blum et~al.}(2015)\citenamefont{Blum, O'Brien, Lauk,
  Bushev, Fleischhauer, and Morigi}}]{blum2015interfacing}
\bibinfo{author}{\bibfnamefont{S.}~\bibnamefont{Blum}},
  \bibinfo{author}{\bibfnamefont{C.}~\bibnamefont{O'Brien}},
  \bibinfo{author}{\bibfnamefont{N.}~\bibnamefont{Lauk}},
  \bibinfo{author}{\bibfnamefont{P.}~\bibnamefont{Bushev}},
  \bibinfo{author}{\bibfnamefont{M.}~\bibnamefont{Fleischhauer}},
  \bibnamefont{and} \bibinfo{author}{\bibfnamefont{G.}~\bibnamefont{Morigi}},
  \bibinfo{journal}{Phys. Rev. A} \textbf{\bibinfo{volume}{91}},
  \bibinfo{pages}{033834} (\bibinfo{year}{2015}).

\bibitem[{\citenamefont{Lai et~al.}(2018)\citenamefont{Lai, Lin, Twamley, and
  Goan}}]{lai2018single}
\bibinfo{author}{\bibfnamefont{Y.-Y.} \bibnamefont{Lai}},
  \bibinfo{author}{\bibfnamefont{G.-D.} \bibnamefont{Lin}},
  \bibinfo{author}{\bibfnamefont{J.}~\bibnamefont{Twamley}}, \bibnamefont{and}
  \bibinfo{author}{\bibfnamefont{H.-S.} \bibnamefont{Goan}},
  \bibinfo{journal}{Phys. Rev. A} \textbf{\bibinfo{volume}{97}},
  \bibinfo{pages}{052303} (\bibinfo{year}{2018}).

\bibitem[{\citenamefont{O'Brien et~al.}(2014)\citenamefont{O'Brien, Lauk, Blum,
  Morigi, and Fleischhauer}}]{o2014interfacing}
\bibinfo{author}{\bibfnamefont{C.}~\bibnamefont{O'Brien}},
  \bibinfo{author}{\bibfnamefont{N.}~\bibnamefont{Lauk}},
  \bibinfo{author}{\bibfnamefont{S.}~\bibnamefont{Blum}},
  \bibinfo{author}{\bibfnamefont{G.}~\bibnamefont{Morigi}}, \bibnamefont{and}
  \bibinfo{author}{\bibfnamefont{M.}~\bibnamefont{Fleischhauer}},
  \bibinfo{journal}{Phys. Rev. Lett.} \textbf{\bibinfo{volume}{113}},
  \bibinfo{pages}{063603} (\bibinfo{year}{2014}).

\bibitem[{\citenamefont{Paauw et~al.}(2009)\citenamefont{Paauw, Fedorov,
  Harmans, and Mooij}}]{paauw2009tuning}
\bibinfo{author}{\bibfnamefont{F.~G.} \bibnamefont{Paauw}},
  \bibinfo{author}{\bibfnamefont{A.}~\bibnamefont{Fedorov}},
  \bibinfo{author}{\bibfnamefont{C.~J. P.~M.} \bibnamefont{Harmans}},
  \bibnamefont{and} \bibinfo{author}{\bibfnamefont{J.~E.} \bibnamefont{Mooij}},
  \bibinfo{journal}{Phys. Rev. Lett.} \textbf{\bibinfo{volume}{102}},
  \bibinfo{pages}{090501} (\bibinfo{year}{2009}).

\bibitem[{\citenamefont{Serfling}(1974)}]{serfling1974probability}
\bibinfo{author}{\bibfnamefont{R.~J.} \bibnamefont{Serfling}},
  \bibinfo{journal}{Ann. Stat.} \textbf{\bibinfo{volume}{2}},
  \bibinfo{pages}{39} (\bibinfo{year}{1974}).

\bibitem[{\citenamefont{Tomamichel and
  Leverrier}(2017)}]{tomamichel2017largely}
\bibinfo{author}{\bibfnamefont{M.}~\bibnamefont{Tomamichel}} \bibnamefont{and}
  \bibinfo{author}{\bibfnamefont{A.}~\bibnamefont{Leverrier}},
  \bibinfo{journal}{Quantum} \textbf{\bibinfo{volume}{1}}, \bibinfo{pages}{14}
  (\bibinfo{year}{2017}).

\bibitem[{\citenamefont{Sugiyama}(2015)}]{sugiyama2015precision}
\bibinfo{author}{\bibfnamefont{T.}~\bibnamefont{Sugiyama}},
  \bibinfo{journal}{Phys. Rev. A} \textbf{\bibinfo{volume}{91}},
  \bibinfo{pages}{042126} (\bibinfo{year}{2015}).

\bibitem[{\citenamefont{Hoeffding}(1963)}]{hoeffding1963probability}
\bibinfo{author}{\bibfnamefont{W.}~\bibnamefont{Hoeffding}},
  \bibinfo{journal}{Journal of the American Statistical Association}
  \textbf{\bibinfo{volume}{58}}, \bibinfo{pages}{13} (\bibinfo{year}{1963}).

\end{thebibliography}

\end{document}